\documentclass[11pt,a4paper]{article}

\pdfoutput=1  

\usepackage{jheppub}  
\usepackage[utf8]{inputenc} 
\graphicspath{{./figures/}} 
\usepackage{caption}
\usepackage{subcaption}

\newcommand{\e}[1]{\mathrm{e}^{#1}} 
\newcommand{\AdS}[1]{AdS$ _{#1} $} 
\let\Re\relax\DeclareMathOperator{\Re}{Re} 
\let\Im\relax\DeclareMathOperator{\Im}{Im} 

\title{Critical behavior of non-hydrodynamic quasinormal modes in a strongly coupled plasma}

\author[a]{Stefano I. Finazzo,}
\author[b]{Romulo Rougemont,}
\author[b]{Maicon Zaniboni,}
\author[b]{Renato Critelli,}
\author[b]{and Jorge Noronha}

\affiliation[a]{Instituto de F\'{i}sica Te\'orica, Universidade do Estado de S\~{a}o Paulo, Rua Dr. Bento T. Ferraz, 271, CEP 01140-070, S\~{a}o Paulo, SP, Brazil}
\affiliation[b]{Instituto de F\'{i}sica, Universidade de S\~{a}o Paulo, Rua do Mat\~{a}o, 1371, Butant\~{a}, CEP 05508-090, S\~{a}o Paulo, SP, Brazil}

\abstract{We study the behavior of quasinormal modes in a top-down holographic dual corresponding to a strongly coupled $\mathcal{N} = 4$ super Yang-Mills plasma charged under a $U(1)$ subgroup of the global $SU(4)$ R-symmetry. In particular, we analyze the spectra of quasinormal modes in the external scalar and vector diffusion channels near the critical point and obtain the behavior of the characteristic equilibration times of the plasma as the system evolves towards the critical point of its phase diagram. Except close to the critical point, we observe that by increasing the chemical potential one generally increases the damping rate of the quasinormal modes, which leads to a reduction of the characteristic equilibration times in the dual strongly coupled plasma. However, as one approaches the critical point the typical equilibration time (as estimated from the lowest non-hydrodynamic quasinormal mode frequency) increases, although remaining finite, while its derivative with respect to the chemical potential diverges with exponent -1/2. We also find a purely imaginary non-hydrodynamical mode in the vector diffusion channel at nonzero chemical potential which dictates the equilibration time in this channel near the critical point.}

\emailAdd{stefano@ift.unesp.br, romulo@if.usp.br, maiconzs@if.usp.br, renato.critelli@usp.br, noronha@if.usp.br}
\arxivnumber{1610.01519}

\begin{document}
\maketitle

\section{Introduction}

\hspace{5 mm} Quasinormal modes (QNM's) are exponentially damped collective excitations \cite{Vishveshwara:1970zz,Davis:1971gg} that define the characteristic behavior of fluctuations of black holes and black branes (for reviews, see \cite{Nollert:1999ji,Kokkotas:1999bd,Berti:2009kk,Konoplya:2011qq}). The spectra of QNM's collectively describe the linear part of the decaying fluctuations of a disturbed black hole, a phenomenon known as ``quasinormal ringing'', which is analogous to the decaying sound emitted by a brass bell when struck by a mallet \cite{Kac:1966xd}. For this reason, QNM's are of great interest to astrophysical and cosmological observations since they describe the ringdown of possible black hole remnants of binary stars and black hole mergers, which were pivotal to the direct detection of gravitational waves earlier this year \cite{Abbott:2016blz,Abbott:2016nmj}.

In the context of the holographic gauge/gravity duality \cite{Maldacena:1997re,Gubser:1998bc,Witten:1998qj,Witten:1998zw}, the QNM's of asymptotically Anti-de Sitter (AdS) spacetimes carry wealthy information about the near equilibrium behavior of the dual strongly interacting quantum field theory (QFT). In fact, the QNM's associated with the fluctuations of a given bulk field are related to the poles of the retarded Green's function of the dual operator in the QFT \cite{Starinets:2002br,Kovtun:2005ev}. These poles describe hydrodynamic and non-hydrodynamic dispersion relations with which one can not only compute hydrodynamic transport coefficients but also derive upper bounds for characteristic equilibration times of the dual QFT plasma \cite{Horowitz:1999jd}. Additionally, non-hydrodynamic modes play an important role in determining the applicability of the hydrodynamic gradient series, as demonstrated by studies in holography \cite{Heller:2013fn,Buchel:2016cbj} and also in kinetic theory \cite{Florkowski:2016zsi,Denicol:2016bjh,Heller:2016rtz}.    

Previous works \cite{Alanen:2011hh,Janik:2015iry,Janik:2016btb} have dealt with QNM's in bottom-up Einstein-dilaton constructions \cite{Gubser:2008ny,Gubser:2008yx,Gursoy:2010fj,Finazzo:2014cna} exhibiting different kinds of phase transitions at zero chemical potential. Additionally, Ref.\ \cite{Buchel:2015saa} investigated the QNM's associated with scalar operators in a top-down $\mathcal{N}=2^*$ non-conformal plasma also at zero chemical potential. On the other hand, in \cite{Rougemont:2015wca} some of us investigated how the QNM's of an external scalar perturbation and, in particular, the equilibration time associated with the imaginary part of the lowest non-hydrodynamical quasinormal frequency, depends on the temperature and baryon chemical potential in a bottom-up, QCD-like Einstein-Maxwell-dilaton model at finite baryon density \cite{Rougemont:2015ona}. In general, through the holographic correspondence, any question regarding the thermalization process in a given strongly coupled gauge theory necessarily involves a study of the QNM's of its gravity dual. These modes describe different timescales in the gauge theory and, close to a critical point, one may expect that the QNM's of the corresponding gravity dual display critical behavior. 
 
Near a critical point, thermodynamical quantities typically display fast variations which enable the definition of critical exponents. Static properties such as single time correlation functions and linear response coefficients to time-independent perturbations display critical behavior which are determined by the underlying equilibrium distribution. However, anomalous behavior is also observed in many dynamical quantities such as the transport coefficients, which depend on the properties of multi-time correlations functions and are not determined by the information contained in the equilibrium distribution. In fact, while static thermodynamical properties of several different physical systems may be grouped into a few different (static) universality classes, dynamical properties associated with slowly varying hydrodynamical fluctuations of a system near criticality do not fit into this static classification scheme, as discussed in detail in \cite{RevModPhys.49.435} nearly 40 years ago. As a matter of fact, the dynamic universality classes reviewed in \cite{RevModPhys.49.435} require the study of hydrodynamic modes, i.e., collective excitations whose frequency vanishes in the case of homogeneous disturbances. While these modes dominate the long time behavior of the system (since they are associated with conserved currents) and can be used to study how transport coefficients (such as the shear viscosity) behave near a critical point, it is conceivable that there is more information about dynamical critical phenomena in multi-time correlation functions that cannot be obtained from their zero frequency limit.

In this paper we initiate the investigation of the critical behavior displayed by non-hydrodynamic modes in strongly coupled gauge theories with gravity duals, i.e., QNM's corresponding to collective excitations in the dual plasma whose frequency does not vanish in the zero wavenumber limit. This novel type of critical phenomena determines the behavior of different characteristic equilibration times of the system at zero wavenumber and, since these QNM's are not directly associated with conserved currents, their behavior at criticality does not follow from the analysis made in Ref.\ \cite{RevModPhys.49.435}. In the present work, as the first exploration in this new arena, we compute QNM's for an external scalar perturbation and also for the diffusion channel associated with a vector perturbation in the so-called 1-R charge black hole (1RCBH) model \cite{Gubser:1998jb,Behrndt:1998jd,Kraus:1998hv,Cai:1998ji,Cvetic:1999ne,Cvetic:1999rb}. This is an analytical top-down construction obtained from $(4+1)$-dimensional maximally supersymmetric gauged supergravity, which is holographically dual to a strongly coupled $\mathcal{N} = 4$ super Yang-Mills (SYM) plasma in flat $(3+1)$ dimensions with a finite chemical potential under a $U(1)$ subgroup of the global $SU(4)$ symmetry of R-charges. This theory is conformal and its phase diagram is a function of a single dimensionless ratio $\mu/T$, where $\mu$ and $T$ are the $U(1)$ R-charge chemical potential and temperature of the black brane background, respectively. The model has a very simple phase diagram with a critical point at $\mu/T = \pi/\sqrt{2}$ and its static critical exponents were computed in \cite{Maeda:2008hn} and \cite{Buchel:2010gd}. Also, the fact that the R-charge conductivity remains finite at the critical point \cite{Maeda:2008hn} shows that this model belongs to the type B dynamical universality class \cite{RevModPhys.49.435} and the anomalous static critical exponent was found to vanish in \cite{Buchel:2010gd}. Thus, this model is of mean-field type \cite{Buchel:2010gd}, which was later argued \cite{Natsuume:2010bs} to be a general consequence of the underlying large $N_c$ approximation. This simple model provides a useful arena for investigating dynamical phenomena in a strongly coupled plasma at finite temperature and density, even though it does not possess the full set of physical properties (such as chiral symmetry) displayed by the real world quark-gluon plasma (QGP) \cite{Rougemont:2016etk}. In fact, such a model may be useful for discovering new dynamical phenomena associated with critical endpoints in strongly coupled non-Abelian plasmas which could be further investigated in more realistic models of the QGP such as \cite{Rougemont:2015wca,Rougemont:2015ona} , with a view towards applications to the ongoing beam energy scan program at RHIC. Other studies of critical phenomena in holography include Refs.\ \cite{Buchel:2009hv,Buchel:2009mf,Buchel:2010ys,DeWolfe:2010he,Cai:2012xh}.

As we are going to show in the next sections, the real and imaginary parts of non-hydrodynamical modes in the external scalar and vector diffusion channels display an infinite slope at the critical point of the phase diagram of the 1RCBH model. This holds true also for higher order QNM's, showing that high frequency modes are also sensitive to the presence of the critical point. In particular, from the imaginary part of the QNM's it is possible to extract the behavior of different characteristic equilibration times in the finite density plasma at criticality (at zero wavenumber) and define a dynamical critical exponent associated with their derivatives with respect to the dimensionless ratio $\mu/T$. We find the same critical exponent $1/2$ for all the equilibration times investigated in the different channels. Except close to the critical point, we observe that by increasing the chemical potential one generally increases the damping of the quasinormal black brane oscillations which, consequently, leads to a reduction of the characteristic equilibration times of the dual plasma. However, as one approaches the critical point these equilibration times are enhanced (though they remain finite) and they acquire an infinite slope. We also find a purely imaginary, non-hydrodynamical mode in the vector diffusion channel at nonzero chemical potential and zero wavenumber which dictates the critical behavior of the equilibration time in this channel (this mode was also found in Ref. \cite{Janiszewski:2015ura} in the context of a $(4+1)$-dimensional Einstein-Maxwell model).

This paper is organized as follows: in Section \ref{sec:1Rcharge} we briefly review the 1RCBH model and its thermodynamics and phase diagram. In Section \ref{sec:qnm} we compute the QNM's of an external scalar perturbation. In Section \ref{sec:vector} we compute the QNM's for the vector diffusion channel in the limit of long wavelengths. Finally, in Section \ref{sec:remarks} we present some final remarks and an outlook. This work is complemented by Appendix \ref{sec:spectral}, where we compute the spectral function in the external scalar channel, and by Appendix \ref{sec:numerics}, where we present a brief discussion regarding the numerical procedures we used for the computation of QNM's and their critical exponents.

Throughout the paper, we work with natural units $\hbar=c=k_B=1$ and use a mostly plus metric signature.

\section{1-R charge black hole model}
\label{sec:1Rcharge}

\hspace{5 mm} For the sake of completeness, in this section we review the thermodynamics of the 1RCBH model \cite{Gubser:1998jb,Behrndt:1998jd,Kraus:1998hv,Cai:1998ji,Cvetic:1999ne,Cvetic:1999rb}. We closely follow the discussion made in Section 4 of Ref.\ \cite{DeWolfe:2011ts} and supplement it with additional plots to better illustrate the behavior of the thermodynamic properties of the model near the critical point. 

\subsection{Background}

\hspace{5 mm} The 1RCBH model is described by an Einstein-Maxwell-dilaton (EMD) action,
\begin{equation}
\label{eq:action}
S = \frac{1}{2 \kappa_5^2} \int d^5x \sqrt{-g} \left[ R - \frac{f(\phi)}{4} F_{\mu \nu} F^{\mu \nu} - \frac{1}{2} (\partial_{\mu} \phi)^2 - V(\phi) \right],
\end{equation}
with the dilaton potential and the coupling between the dilaton and Maxwell fields given by,
\begin{align}
V(\phi) = -\frac{1}{L^2} \left(8\e{\frac{\phi}{\sqrt{6}}} + 4 \e{-\sqrt{\frac{2}{3}}\phi} \right), \,\,\,
f(\phi) = \e{- 2\sqrt{\frac{2}{3}}\phi},
\end{align}
where $\kappa_5^2$ is the five dimensional Newton's constant and $L$ is the asymptotic $\mathrm{AdS}_5$ radius. With these profiles for $V(\phi)$ and $f(\phi)$, one obtains a consistent truncation of maximally supersymmetric gauged supergravity in five dimensions, which is itself a consistent truncation of type IIB superstring theory on AdS$_5 \times S^5$. This model is a \emph{bona fide} top-down string theory construction, which is dual to a SYM plasma with a finite chemical potential under a $U(1)$ subgroup of the global $SU(4)$ symmetry of R-charges. For simplicity, in the following we set $L=1$.

The 1RCBH solution is defined by
\begin{align}
d s^2 & = \e{2A(r)}\left(-h(r) d t^2 +d\vec{x}^2\right) + \frac{\e{2B(r)}}{h(r)} d r^2,\label{eq:line element}\\
A(r) & = \ln  r  + \frac{1}{6} \ln \left( 1 + \frac{Q^2}{r^2} \right), \\
B(r) & = - \ln r  -  \frac{1}{3} \ln \left( 1 + \frac{Q^2}{r^2} \right), \\
h(r) & = 1 - \frac{M^2}{r^2(r^2+Q^2)},\label{eq:metric components} \\
\phi (r) & = -\sqrt{\frac{2}{3}} \ln \left( 1 + \frac{Q^2}{r^2} \right),\label{eq:scalar field} \\
\mathbf{A} & = \Phi(r) dt = \left(-\frac{M Q}{r^2+Q^2} +\frac{M Q}{r_H^2+Q^2}\right)dt,\label{eq:electromagnetic four-potential}
\end{align}
where $r$ is the holographic radial coordinate and the boundary of the asymptotically \AdS{5} geometry is located at $r\to\infty$. The radial position of the black brane horizon may be written in terms of the charge $Q$ and mass $M$ of the black brane as follows,
\begin{equation}
r_H = \sqrt{\frac{\sqrt{Q^4 + 4 M^2} - Q^2}{2}}.
\end{equation}
The 1RCBH background is, thus, characterized by two nonnegative parameters, $(Q,M)$ or, alternatively, $(Q,r_H)$.

The Hawking temperature of the black brane horizon is given by
\begin{align}
\label{eq:temperature}
T =\left.\dfrac{\sqrt{-(g_{tt})'(g^{rr})'}}{4\pi}\right|_{r=r_H} = \frac{Q^2 + 2 r_H^2}{2 \pi \sqrt{Q^2+r_H^2}},
\end{align}
where $'$ denotes a derivative with respect to $r$. On the other hand, the $U(1)$ R-charge chemical potential reads
\begin{align}
\label{eq:chemical potential}
\mu = \lim\limits_{r\to\infty}\Phi(r) = \frac{Q r_H}{\sqrt{Q^2 + r_H^2}}.
\end{align}

\subsection{Phase diagram}

\hspace{5 mm} The class of solutions corresponding to the 1RCBH model may be parametrized by different values of the dimensionless ratio $Q/r_H$. By dividing Eqs.\ \eqref{eq:temperature} and \eqref{eq:chemical potential}, and then solving for $Q/r_H$, one obtains
\begin{align}\label{eq:critical_point}
\dfrac{Q}{r_H}=\sqrt{2}\left(\dfrac{1\pm\sqrt{1-\left(\frac{\mu/T}{\pi/\sqrt{2}}\right)^2}}{\frac{\mu/T}{\pi/\sqrt{2}}}\right).
\end{align}
Since $Q/r_H$ is nonnegative, \eqref{eq:critical_point} implies that $\mu/T\in\big[0,\pi/\sqrt{2}\big]$. It also follows from \eqref{eq:critical_point} that for every value of $\mu/T\in\big[0,\pi/\sqrt{2}\big)$, there are two different corresponding values of $Q/r_H$, which parametrize two different branches of solutions. As we are going to show in the next subsection, by analyzing the thermodynamics of the 1RCBH solutions one concludes that the point of the phase diagram where these two branches merge, $\mu/T=\pi/\sqrt{2}$ or, correspondingly, $Q/r_H=\sqrt{2}$, is a critical point of a second order phase transition. 

In order to simplify the equations in this paper we define below a useful variable that smoothly connects the two branches of solutions,
\begin{align} \label{eq:y and mu/T}
y^2+\left(\frac{\mu/T}{\pi/\sqrt{2}}\right)^2=1\quad\text{ with }\quad y\in[-1,1],
\end{align}
where $y=0$ parametrizes the critical background geometry with $\mu/T=\pi/\sqrt{2}$, while $y=1$ parametrizes the \AdS{5}-Schwarzschild background with zero charge, corresponding to $Q=0$ and $r_H\neq 0$, which implies $\mu/T=0$. For $y=-1$ we also have $\mu/T=0$, but this time $r_H=0$ and $Q\neq 0$, which corresponds to a supersymmetric BPS solution dubbed ``superstar'' \cite{Myers:2001aq} instead of a black hole.

\begin{figure}[t]
\centering
\begin{subfigure}{0.49\textwidth}
\includegraphics[width=\textwidth]{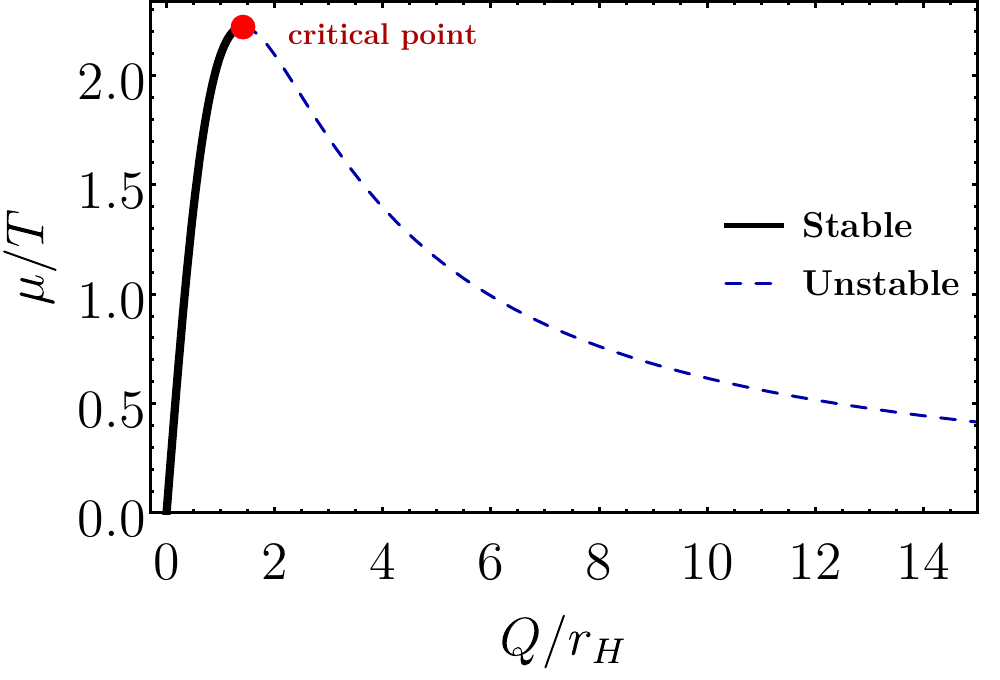}
\caption{}
\end{subfigure}
\begin{subfigure}{0.49\textwidth}
\includegraphics[width=\textwidth]{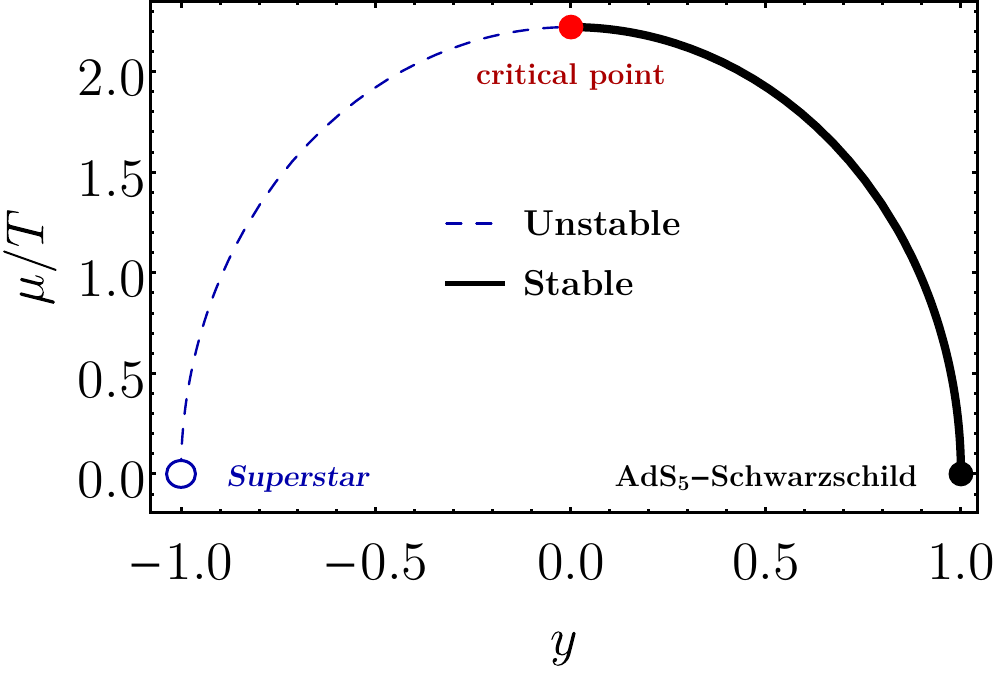}
\caption{}
\end{subfigure}
\caption{(Color online) Phase structure of the 1RCBH model (closely following the discussion in Ref. \cite{DeWolfe:2011ts}): (a) the single dimensionless control parameter of the QFT phase diagram, $\mu/T$, as a function of the corresponding dimensionless ratio $Q/r_H$ on the gravity side for both stable and unstable branches (note that the superstar solution lies at $Q/r_H\to\infty$); (b) the same, now in terms of the alternative variable $y$ defined in Eq.\ \eqref{eq:y and mu/T}.}
\label{fig:critical point}
\end{figure}

As we are going to see in the next subsection, the thermodynamically stable branch corresponds to the lower sign in Eq.\ \eqref{eq:critical_point} with $Q/r_H \in \big[0,\sqrt{2}\big)$ or $y\in \big(0,1\big]$, while the thermodynamically unstable branch corresponds to the upper sign in Eq.\ \eqref{eq:critical_point} with $Q/r_H\in\big(\sqrt{2},\infty\big)$ or $y\in\big[-1,0\big)$, with both branches of solutions being smoothly connected at the critical point, $Q/r_H=\sqrt{2}$ or $y=0$. This is illustrated in Fig.\ \ref{fig:critical point}.

\subsection{Thermodynamics: equation of state and susceptibilities}

\hspace{5 mm} For a SYM plasma, it is known \cite{Gubser:1996de} that
\begin{equation}
\frac{1}{\kappa_5^2} = \frac{N_c^2}{4\pi^2}.
\end{equation}
By substituting the expression above in Bekenstein-Hawking's relation \cite{Bekenstein:1973ur,Hawking:1974sw}, one can write down the entropy density as follows
\begin{align}\label{eq:entropy density}
\dfrac{s}{N_c^2T^3}=\dfrac{\pi^2}{16}\left(3\pm\sqrt{1-\left(\dfrac{\mu/T}{\pi/\sqrt{2}}\right)^2}\right)^2 \left(1\mp\sqrt{1-\left(\dfrac{\mu/T}{\pi/\sqrt{2}}\right)^2}\right).
\end{align}
The R-charge density, $\rho=\lim_{r\to\infty}\delta S/\delta\Phi'$, may be written as
\begin{align}\label{eq:charge density}
\dfrac{\rho}{N_c^2T^3}=\dfrac{\mu/T}{16}\left(3\pm\sqrt{1-\left(\dfrac{\mu/T}{\pi/\sqrt{2}}\right)^2}\right)^2,
\end{align}
where the lower/upper signs denote the stable/unstable branches, as in Eq.\ \eqref{eq:critical_point} (we note that in this model any dimensionless ratio is always written in terms of $\mu/T$, as expected).

From the Gibbs-Duhem relation, $dp=sdT+\rho d\mu$, one may compute the pressure,
\begin{align}\label{eq:pressure}
\dfrac{p}{N_c^2 T^4}= \dfrac{\pi^2}{128}\left(3\pm\sqrt{1-\left(\dfrac{\mu/T}{\pi/\sqrt{2}}\right)^2}\right)^3 \left(1\mp\sqrt{1-\left(\dfrac{\mu/T}{\pi/\sqrt{2}}\right)^2}\right).
\end{align}
Using Eqs.\ \eqref{eq:entropy density}, \eqref{eq:charge density}, and \eqref{eq:pressure} one can readily evaluate the internal energy density, $ \varepsilon=Ts-p+\mu\rho$, obtaining $\varepsilon=3p$, as expected for a conformal QFT in four dimensions.

\begin{figure}[t]
	\centering
	\begin{subfigure}{0.49\textwidth}
		\includegraphics[width=\textwidth]{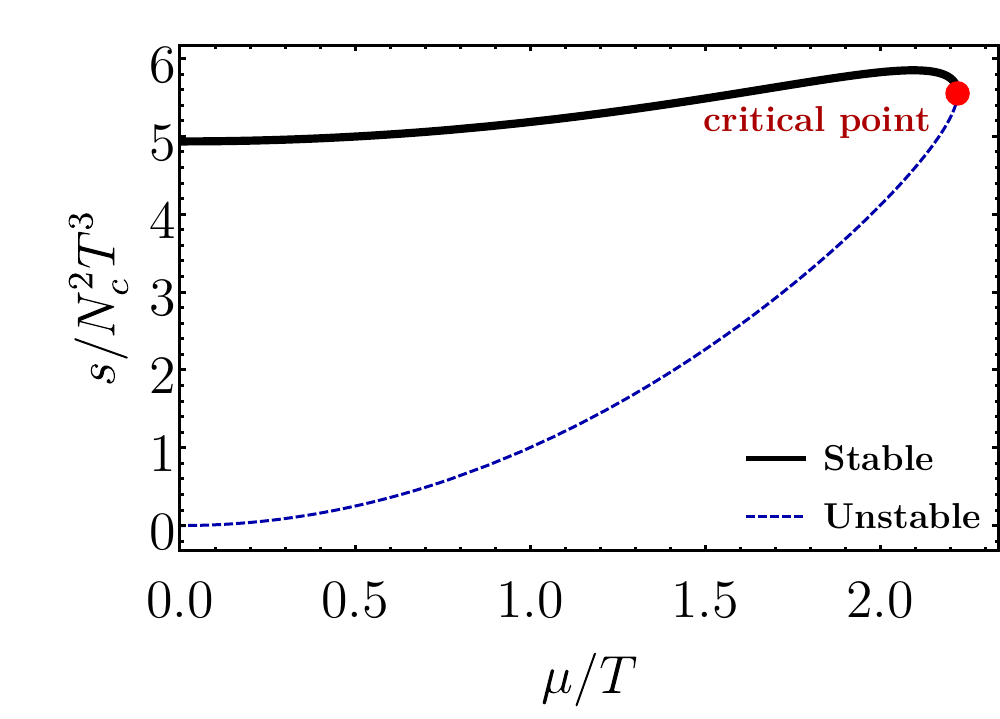}
		\caption{Entropy density}
	\end{subfigure}
	\begin{subfigure}{0.49\textwidth}
		\includegraphics[width=\textwidth]{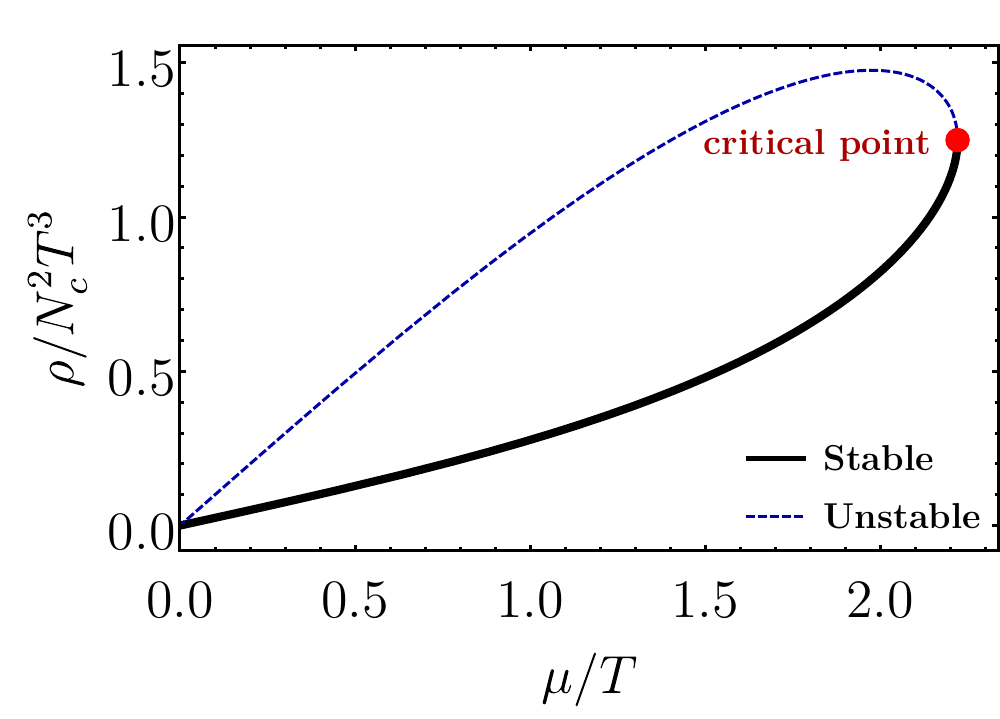}
		\caption{Charge density}
	\end{subfigure}
	\begin{subfigure}{0.49\textwidth}
		\includegraphics[width=\textwidth]{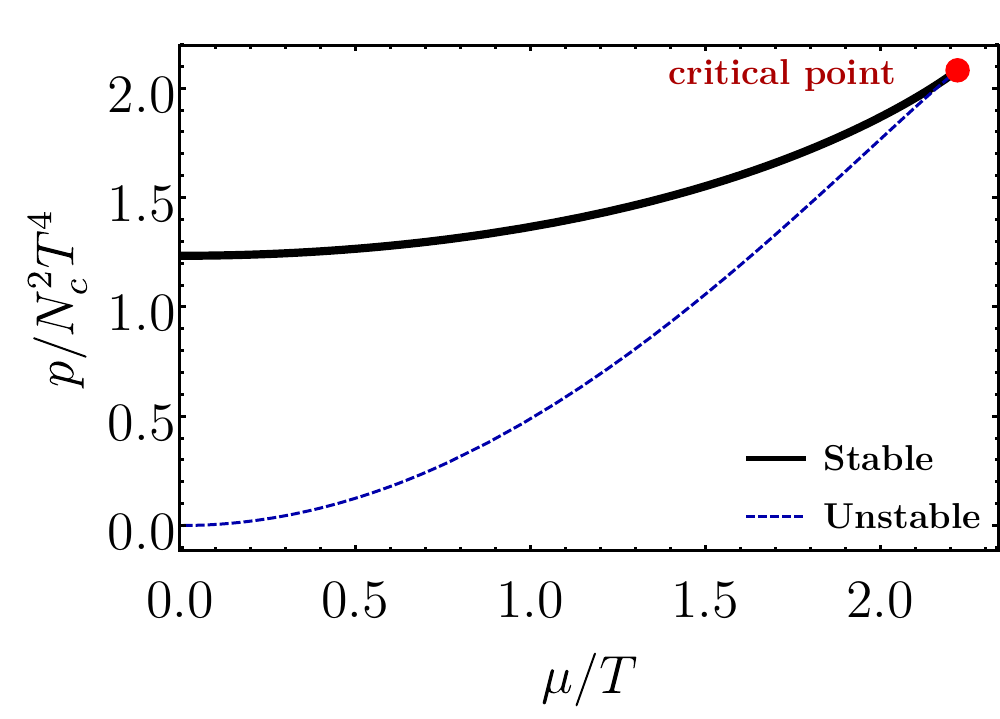}
		\caption{Pressure}
	\end{subfigure}
	\begin{subfigure}{0.49\textwidth}
		\includegraphics[width=\textwidth]{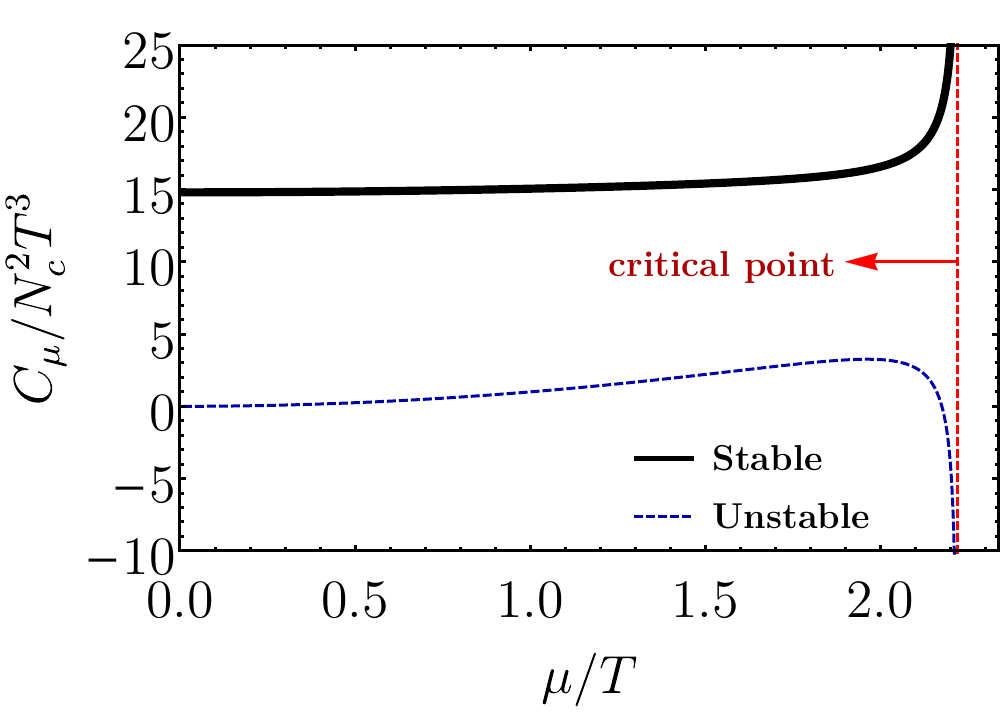}
		\caption{Heat capacity}
	\end{subfigure}
	\caption{(Color online) Equation of state and heat capacity for the 1RCBH model(closely following the discussion in Ref. \cite{DeWolfe:2011ts}).} \label{fig:thermo}
\end{figure}

\begin{figure}[t]
\centering
\begin{subfigure}{0.49\textwidth}
	\includegraphics[width=\textwidth]{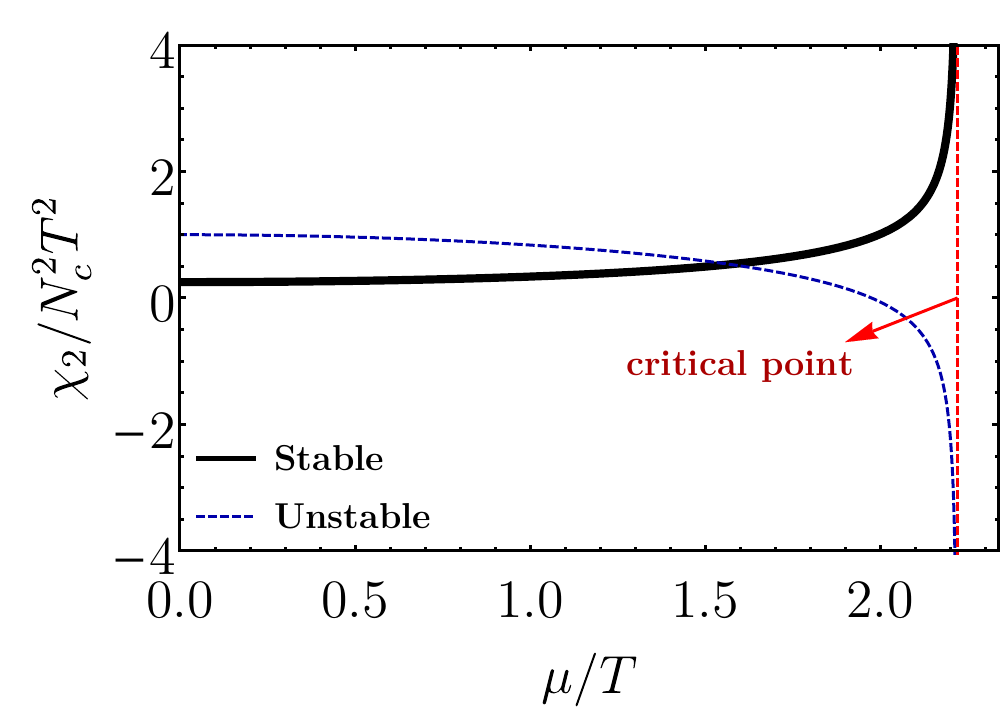}
	\caption{2nd-order susceptibility}
\end{subfigure}
\begin{subfigure}{0.49\textwidth}
	\includegraphics[width=\textwidth]{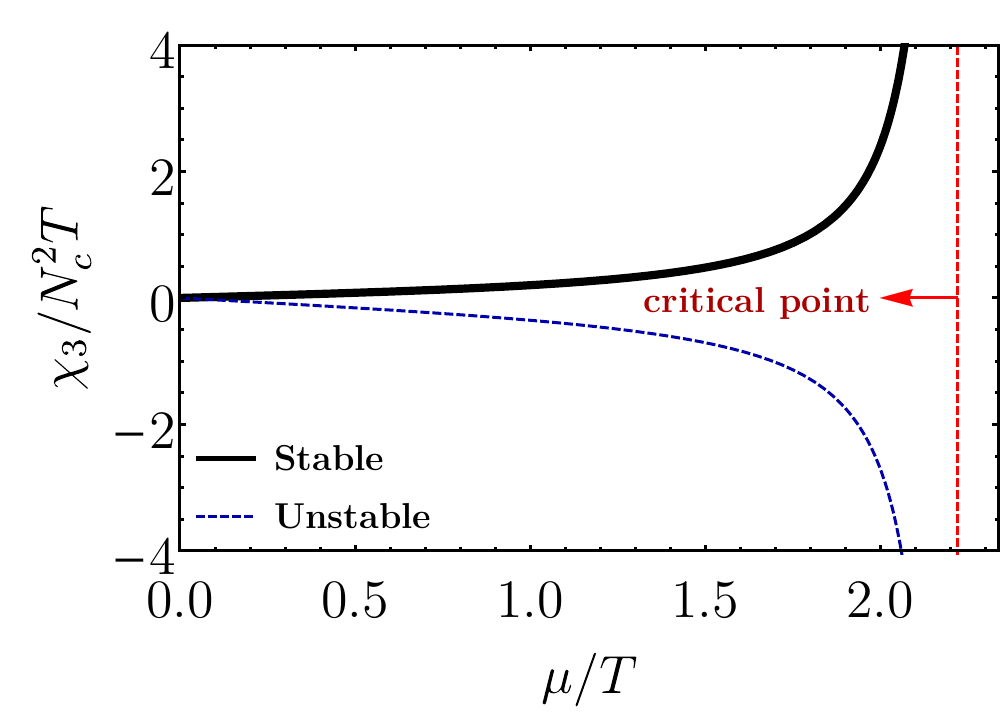}
	\caption{3rd-order susceptibility}
\end{subfigure}
\begin{subfigure}{0.49\textwidth}
	\includegraphics[width=\textwidth]{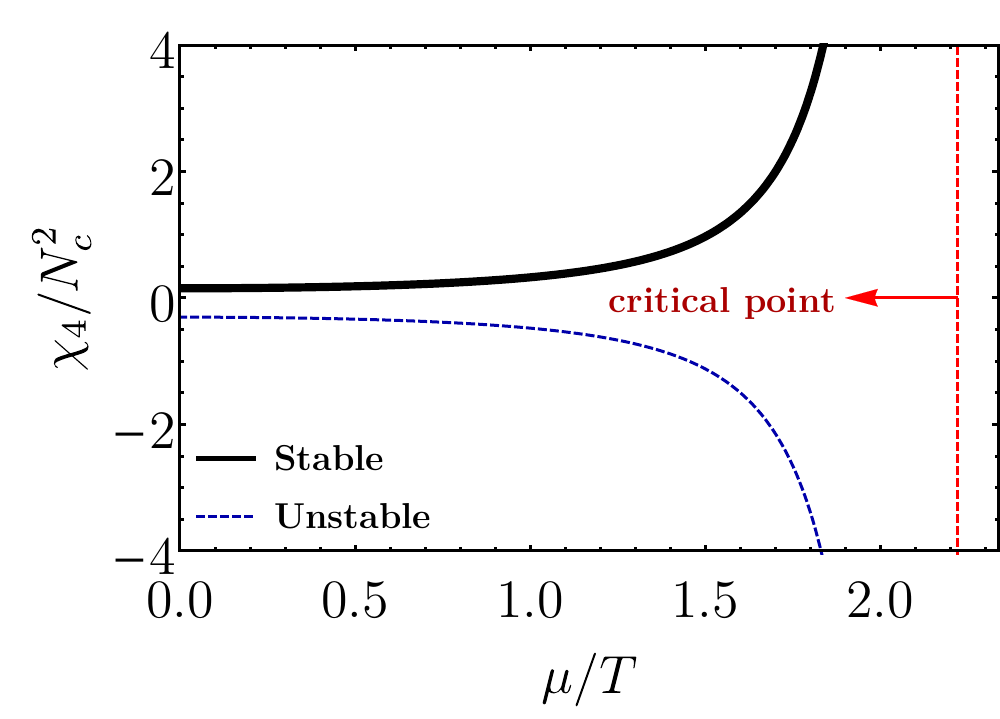}
	\caption{4th-order susceptibility}
\end{subfigure}
\begin{subfigure}{0.49\textwidth}
	\includegraphics[width=\textwidth]{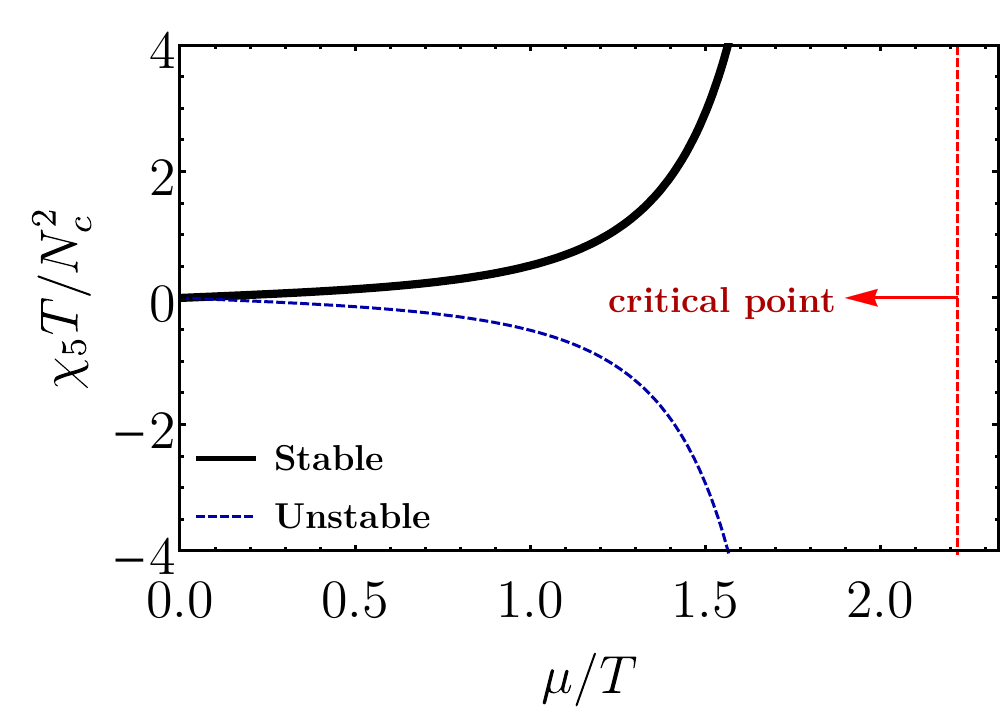}
	\caption{5th-order susceptibility}
\end{subfigure}
\begin{subfigure}{0.49\textwidth}
	\includegraphics[width=\textwidth]{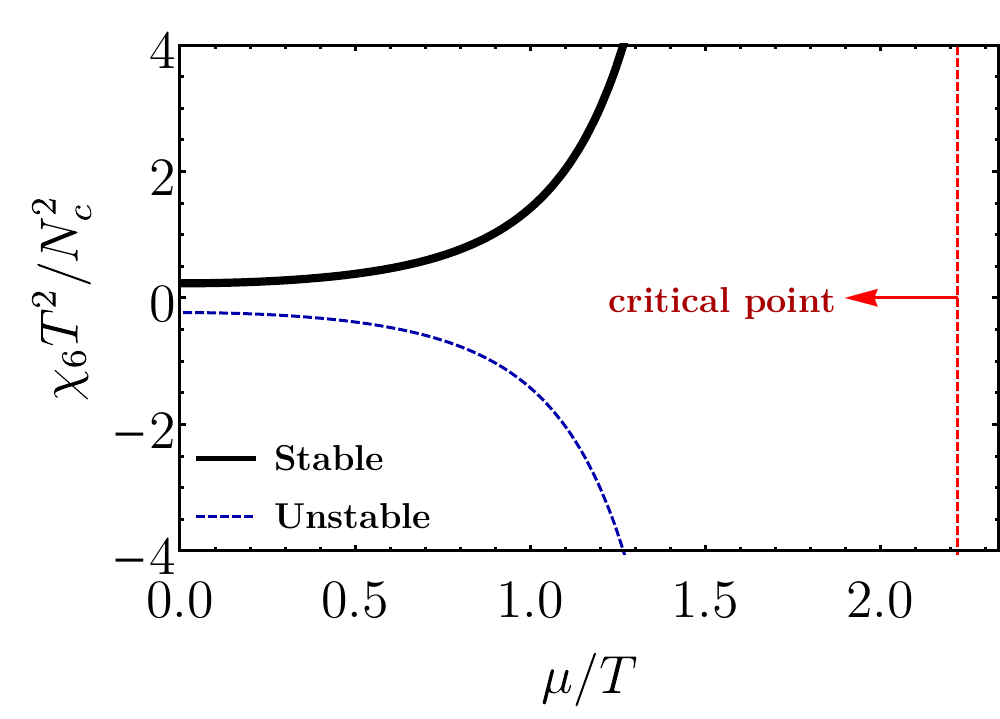}
	\caption{6th-order susceptibility}
\end{subfigure}
\begin{subfigure}{0.49\textwidth}
	\includegraphics[width=\textwidth]{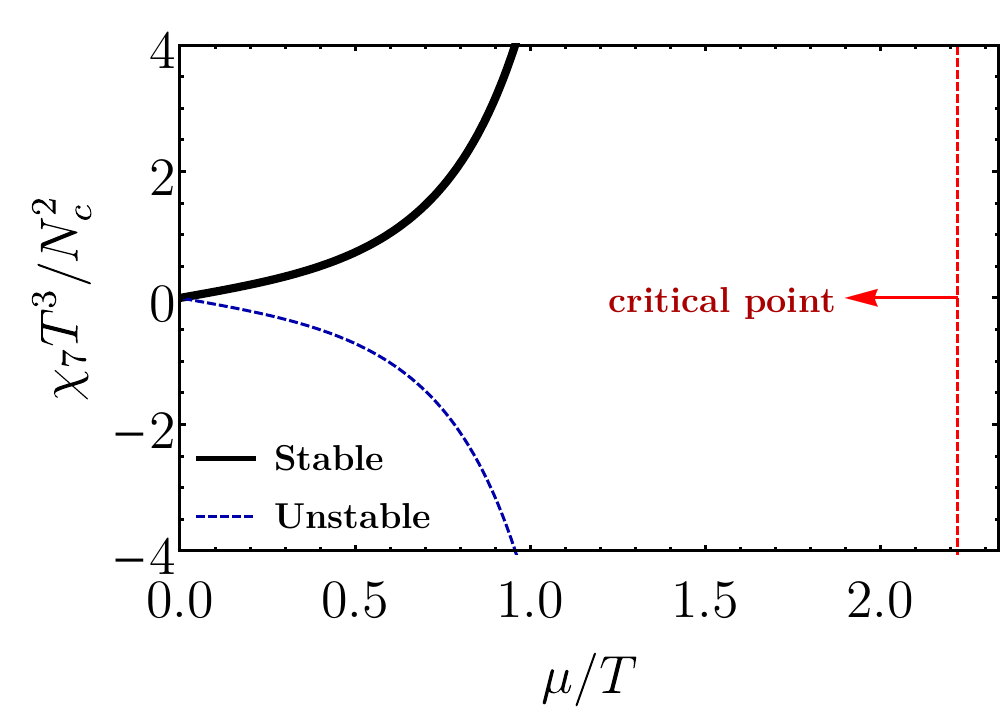}
	\caption{7th-order susceptibility}
\end{subfigure}
\caption{(Color online) $n$th-order susceptibilities for the 1RCBH model.} \label{fig:susceptibility}
\end{figure}

The heat capacity at fixed chemical potential is given by $C_\mu=\left(\partial s/\partial T\right)_\mu$, while the $n$th-order R-charge susceptibility is given by $\chi_n=\left(\partial^n p/\partial \mu^n\right)_T$ (note that $\chi_1=\rho$).
At the critical point the heat capacity $C_\mu$ and the higher order susceptibilities $\chi_{n\ge 2}$ diverge. In Fig.\ \ref{fig:thermo} we show the equation of state and the heat capacity while in Fig.\ \ref{fig:susceptibility} we display the susceptibilities for the stable and unstable branches of the 1RCBH model.

Thermodynamical stability is ensured if the Jacobian $\mathcal{J} = \partial(s,\rho)/\partial(T,\mu)$ is positive. In the 1RCBH model one obtains
\begin{align}\label{eq:Jacobian}
\dfrac{\mathcal{J}}{N_c^4T^4}=\dfrac{3\pi^2}{256}(3-y)^4\left(1+\dfrac{1}{y}\right).
\end{align}
In the equation above, the quartic term is always positive while the expression in the last parenthesis becomes negative for $y\in\big(-1,0\big)$, justifying the aforementioned classification of stable and unstable branches.\footnote{Note that the superstar solution ($y=-1$) has $\mathcal{J}=0$ and corresponds to a saddle point.}

Finally, by analyzing the behavior of the R-charge density in \eqref{eq:charge density} near the critical point one can see that the static critical exponent $\delta=2$, as discussed in \cite{Maeda:2008hn,Buchel:2010gd,DeWolfe:2011ts}.

\section{QNM's for an external scalar fluctuation}
\label{sec:qnm}

\subsection{Equation of motion}

\hspace{5 mm} In the previous section we reviewed the thermodynamical equilibrium properties of the 1RCBH plasma. We now analyze near-equilibrium properties of the system encoded in its quasinormal modes. In this section we calculate the QNM's for an external scalar perturbation $\varphi$ on top of the 1RCBH backgrounds, which is described by the bulk action,
\begin{equation}
\label{eq:scalaraction}
S = \frac{1}{2\kappa_5^2} \int d^5 x \sqrt{-g} \left[-\frac{1}{2}(\partial_\mu\varphi)^2\right].
\end{equation}

The equation of motion following from this action is just the massless Klein-Gordon equation on top of the solution given by Eq.\ \eqref{eq:line element}. We take a plane-wave Ansatz for the Fourier modes of the perturbation, $\varphi=\e{-i\omega t+i\vec{k}\cdot\vec{x}}\tilde{\varphi}(\omega,\vec{k},r)$, which for brevity we write simply as $\tilde{\varphi}(\omega,\vec{k},r)\equiv\tilde{\varphi}(r)$. The resulting equation of motion then only depends on the frequency $\omega$, the magnitude of the spatial 3-momentum $k\equiv |\vec{k}|$, and the background control parameter $y$.

In what follows we employ the in-falling Eddington-Finkelstein (EF) ``time'' coordinate defined by,
\begin{equation}
d v = d t + \sqrt{-\frac{g_{rr}}{g_{tt}}}d r = d t + \frac{\e{B-A}}{h}dr,
\end{equation}
in terms of which the metric \eqref{eq:line element} becomes
\begin{equation}
d s^2 = \e{2A}\left(-hd v^2+d\vec{x}^2\right) + 2 \e{A+B} dv dr.
\end{equation}
One of the main advantages of the EF coordinates is that the in-falling wave condition at the horizon, which is associated with the retarded Green's function, becomes automatically satisfied by just requiring regularity of the solutions there. In these coordinates, the equation of motion for $\tilde{\varphi}$ becomes,
\begin{align}
\tilde{\varphi}''+\left(4A'-B'+\dfrac{h'}{h}\right)\tilde{\varphi}'-i\omega\dfrac{\e{B-A}}{h}(2\tilde{\varphi}' +3A'\tilde{\varphi})-k^2\dfrac{\e{2(B-A)}}{h}\tilde{\varphi}=0.
\end{align}
We map the radial coordinate $r$, defined on the interval $r_H \le r < \infty$, to a new dimensionless radial coordinate $u = r_H/r$, defined on the interval $0 \le u \le 1$, which is more suitable to be used in the pseudospectral method \cite{boyd01} (to be briefly reviewed in Appendix \ref{sec:numerics}). In these new coordinates, the equation of motion for the external scalar perturbation becomes,
\begin{align}
& \tilde{\varphi} ''-\frac{\left(u^4 (3-y)+2 u^2 (1-y)-3 (1+y)\right) \tilde{\varphi} '}{u \left(1-u^2\right) \left(u^2 (3-y)+1+y\right)}+\nonumber\\
&-\dfrac{2i(\omega/T)}{\pi  \left(1-u^2\right) \sqrt{3-y} \left(u^2 (3-y)+1+y\right)}\left(\frac{\left(4 u^2 (1-y)+3 (y+1)\right)\tilde{\varphi} }{u \sqrt{2 u^2 (1-y)+1+y}}+\right.\nonumber\\
&+2 \sqrt{2 u^2 (1-y)+1+y}\, \tilde{\varphi} '\Bigg)-\frac{4 (k/T)^2 \tilde{\varphi} }{\pi ^2 \left(1-u^2\right) (3-y) \left(u^2 (3-y)+1+y\right)}=0,
\end{align}
where the primes now denote derivatives with respect to the new radial coordinate $u$. From the discussion above, and from the definition of the background control parameter $y$ in Eq. \eqref{eq:y and mu/T}, one concludes that the dimensionless quasinormal eigenfrequencies, $\omega/T$, will depend only on the dimensionless ratios $\mu/T$ and $k/T$.

To completely specify the eigenvalue problem to be solved in order to find the QNM spectra associated to this external scalar perturbation, we still need to impose a Dirichlet boundary condition. From the fact that $\tilde{\varphi}$ is a scalar field defined on an asymptotically \AdS{5} background, it follows that asymptotically close to the boundary it may be written as $\tilde{\varphi} (u) = G(u) + u^4 F(u)$, with the leading, non-normalizable mode $G(u\to 0)=J(\omega,\vec{k})$ being the source for the QFT operator $\hat{O}$ dual to the (external) scalar field $\varphi$, and the subleading, normalizable mode $F(u\to 0)=\langle\hat{O}(\omega,\vec{k})\rangle$ being its expectation value. According to the real time holographic dictionary \cite{Son:2002sd}, the retarded propagator of the QFT operator $\hat{O}$ is given the ratio between the normalizable and non-normalizable modes, $\mathcal{G}^R_{\hat{O}\hat{O}}(\omega,\vec{k})=-\langle\hat{O}(\omega,\vec{k})\rangle/J(\omega,\vec{k})$, therefore, if we impose as a Dirichlet boundary condition the selection of the normalizable mode by setting $G(0)=0$ with $F(0)\neq 0$, we are left with an eigenvalue problem whose eigenfrequencies correspond to dispersion relations $\omega/T=\omega(k/T;\mu/T)/T$ describing the poles of $\mathcal{G}^R_{\hat{O}\hat{O}}$, which are the QNM's we are looking for. Then, we set $\tilde{\varphi} (u) = u^4 F(u)$, with $F(0)\neq 0$, from which it follows that,
\begin{align}
\label{eq:qnmfinal}
& 16 u \left(1-\frac{2}{u^2 (3-y)+1+y}\right)F+\left(u^2 \left(9-\frac{8}{u^2 (3-y)+1+y}\right)-5\right) F'+\nonumber\\
&-u \left(1-u^2\right) F''+\dfrac{2i(\omega/T)}{\pi \sqrt{3-y} \left(u^2 (3-y)+1+y\right)} \left(-\frac{\left(12 u^2 (1-y)+5 (1+y)\right)F}{\sqrt{2 u^2 (1-y)+1+y}}+\right.\nonumber\\
&+2 u \sqrt{2 u^2 (1-y)+1+y}\, F'\Bigg)+\frac{4 (k/T)^2 u F}{\pi ^2 (3-y) \left(u^2 (3-y)+1+y\right)}=0.
\end{align}
We now have a Generalized Eigenvalue Problem (GEP) for the eigenfunction $F(u)$ and the quasinormal eigenfrequency $\omega/T$, which may be solved as functions of $k/T$ and $\mu/T$. In this work we use the pseudospectral method.

Note that Eq.\ \eqref{eq:qnmfinal} also reveals one of the greatest virtues of the EF coordinates, namely, the fact that it reduces the QNM eigenvalue problem from a Quadratic Eigenvalue Problem \cite{QEP} in the standard set of spacetime coordinates to a GEP, which requires far less computational cost when numerically evaluating the QNM spectra.

\subsection{QNM spectra and equilibration time}
\label{sec:QNMexternal}

\begin{figure}[t]
\centering
\includegraphics[width=0.7\textwidth]{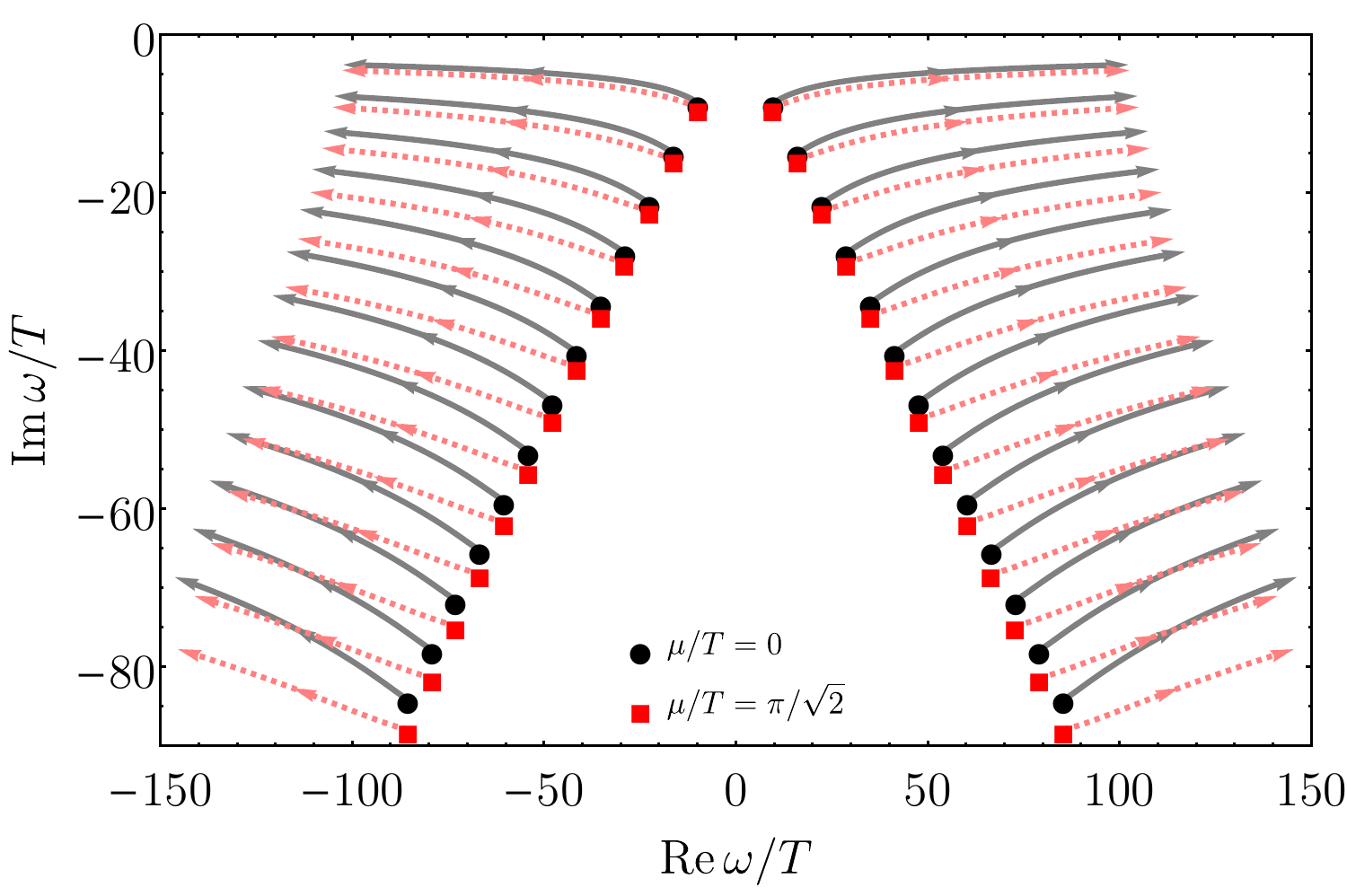}
\caption{(Color online) First 26 QNM's trajectories in the external scalar channel evolved within the interval $0\le k/T\leq 100$ for $\mu/T = 0$ (beginning with a black dot for $ k=0 $ and evolving into solid gray lines for $ k>0 $) and for the critical point $\mu/T = \pi/\sqrt 2$ (beginning with a red square for $ k=0 $ and evolving into dashed pink lines for $ k>0 $).}
\label{fig:qnmtyp}
\end{figure}

\hspace{5 mm} In Fig.\ \ref{fig:qnmtyp} we show the evolution of the external scalar QNM spectra for the first 26 poles as we evolve $k/T$ from 0 to 100, both for the \AdS{5}-Schwarzschild background at $\mu/T=0$ and the critical geometry at $\mu/T = \pi/\sqrt{2}$. We observe the usual non-hydrodynamical QNM structure for the external scalar channel with an infinite series of QNM pairs with $\Im\omega < 0$ and $\Re\omega \neq 0$ symmetrically distributed with respect to the imaginary axis \cite{Kovtun:2005ev}. When $k/T=0$, by increasing the $\mu/T$ ratio one increases the magnitude of the imaginary part of the QNM's, which becomes more appreciable for higher order, faster varying modes. On the other hand, an increase in $k/T$ enhances (suppresses) the magnitude of the real (imaginary) part of the poles. We see that by increasing the chemical potential one generally increases the damping of the quasinormal black hole oscillations, which qualitatively agrees with the result found previously in \cite{Rougemont:2015wca} for a non-conformal, QCD-like bottom-up EMD model describing the physics of the QGP at finite baryon density. 

Also, we note that the non-hydrodynamic modes in Fig.\ \ref{fig:qnmtyp} remain finite when evaluated at the critical point, even when $k=0$. Thus, one can see that the timescales contained in the non-hydrodynamic modes are different than the usual relaxation time quantity $\tau_{rel} \sim \xi^z$, where $\xi$ is the correlation length (which diverges at the critical point) and $z$ is the dynamical critical exponent, which becomes infinitely large at criticality describing the well-known phenomenon of critical slowing down. Nevertheless, in this strongly coupled model the microscopic scales defined by the non-hydrodynamic QNM's still display some critical behavior, as we show below.

\begin{figure}[t]
\centering
\begin{subfigure}{0.49\textwidth}	
	\includegraphics[width=\textwidth]{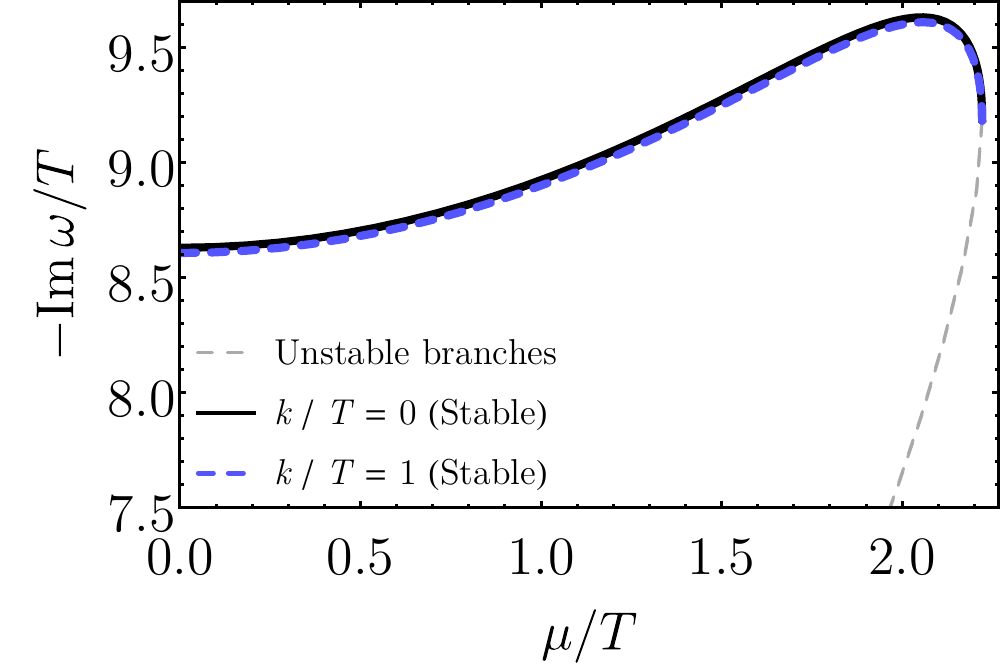}
	\caption{1st QNM}
\end{subfigure}
\begin{subfigure}{0.49\textwidth}	
	\includegraphics[width=\textwidth]{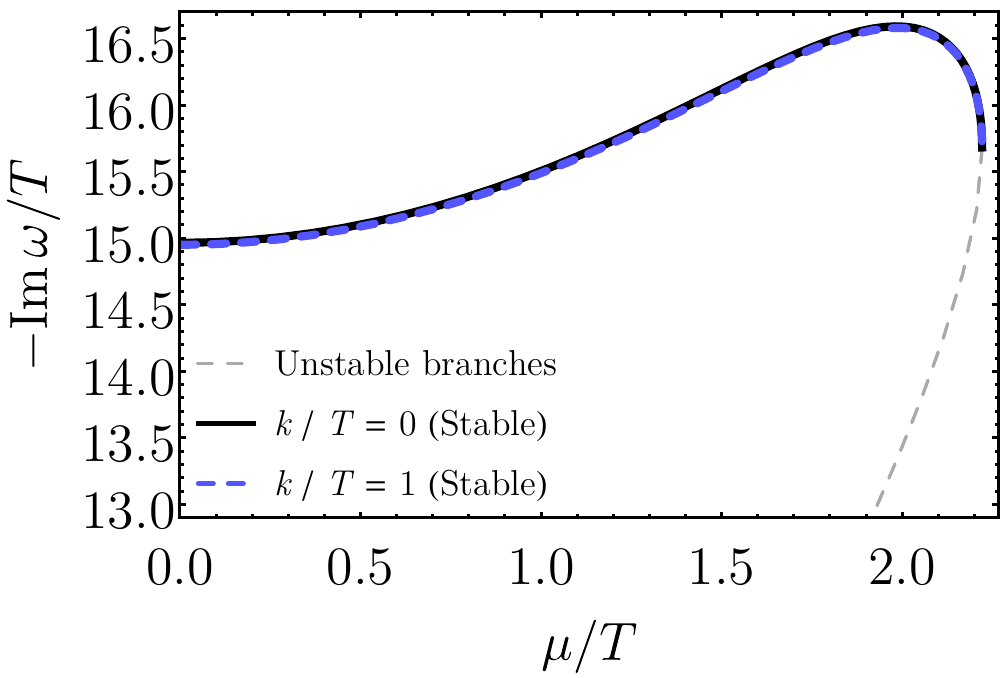}
	\caption{2nd QNM}
\end{subfigure}
\begin{subfigure}{0.49\textwidth}	
	\includegraphics[width=\textwidth]{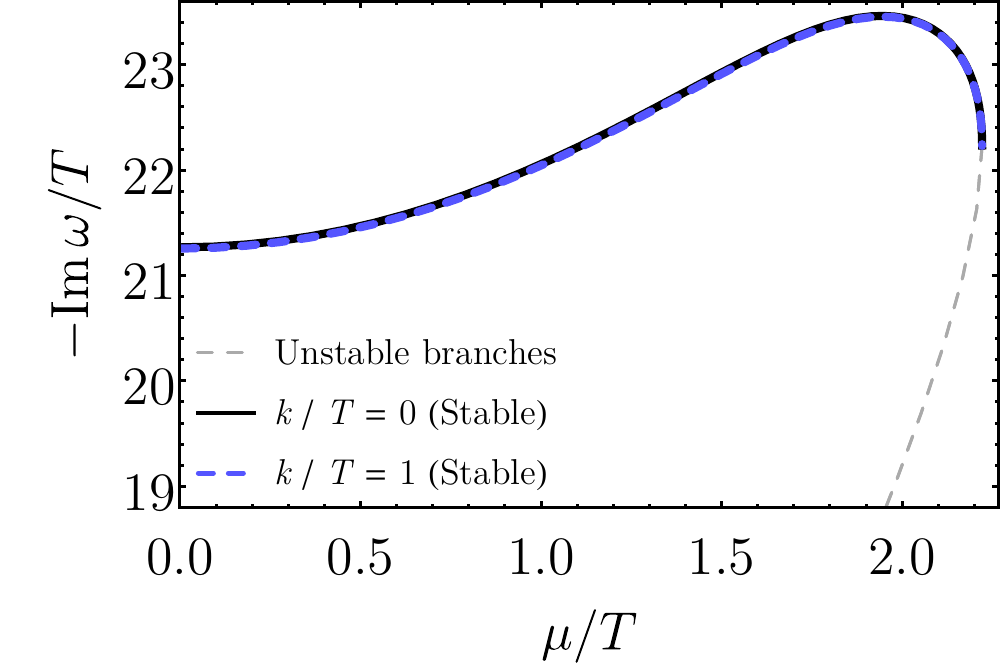}
	\caption{3rd QNM}
\end{subfigure}
\begin{subfigure}{0.49\textwidth}	
	\includegraphics[width=\textwidth]{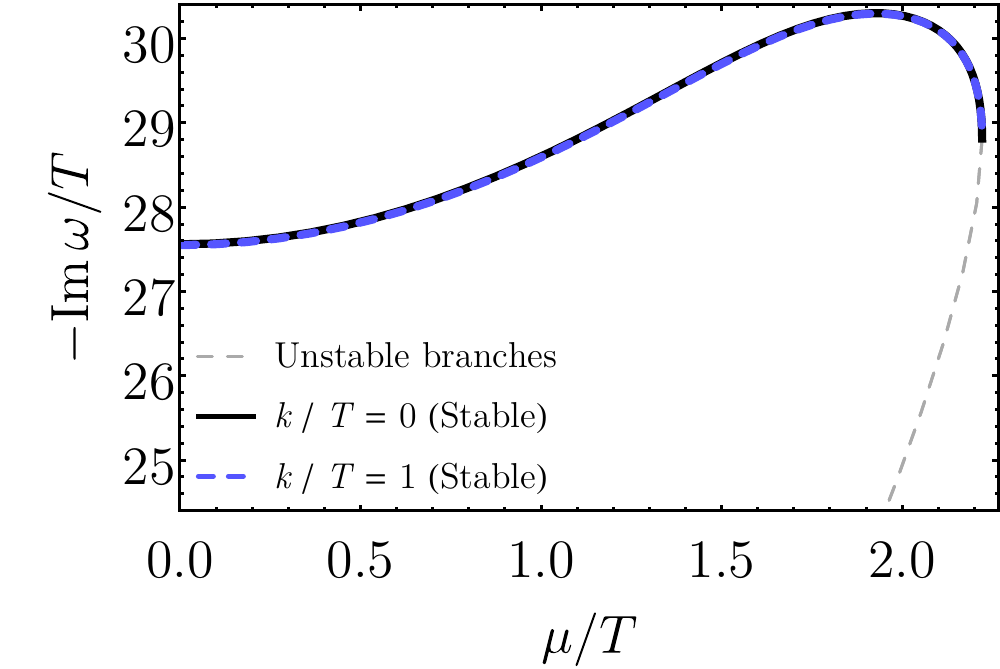}
	\caption{4th QNM}
\end{subfigure}
\caption{(Color online) Imaginary part of the first 4 QNM's in the external scalar channel for $k/T = 0$ and $k/T = 1$, as a function $\mu/T$, for both stable and unstable branches.}
\label{fig:qnmim}
\end{figure}

\begin{figure}[t]
\centering
\begin{subfigure}{0.49\textwidth}	
	\includegraphics[width=\textwidth]{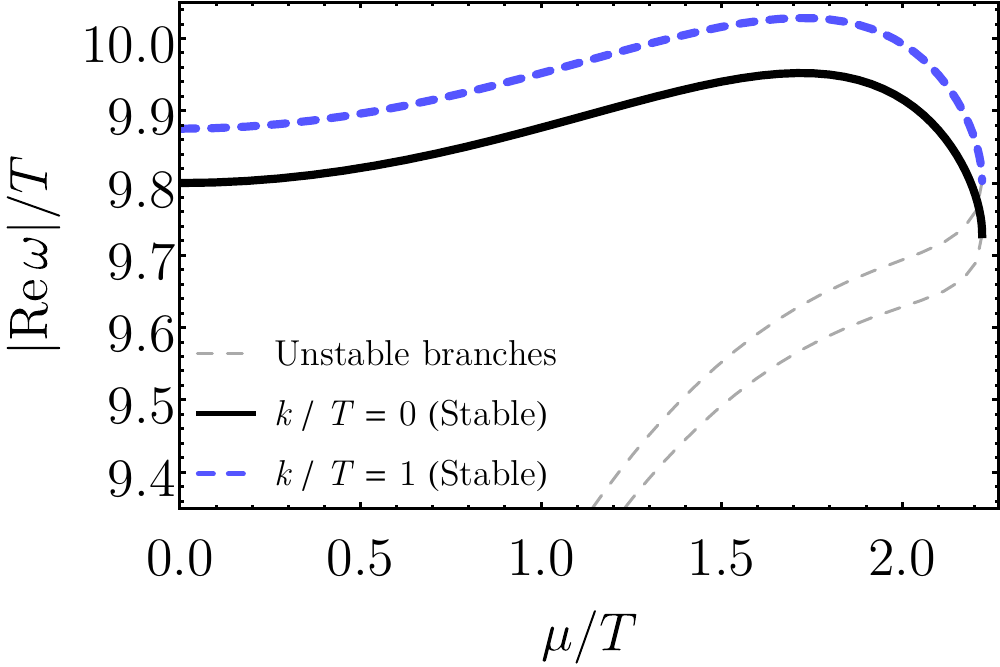}
	\caption{1st QNM}
\end{subfigure}
\begin{subfigure}{0.49\textwidth}	
	\includegraphics[width=\textwidth]{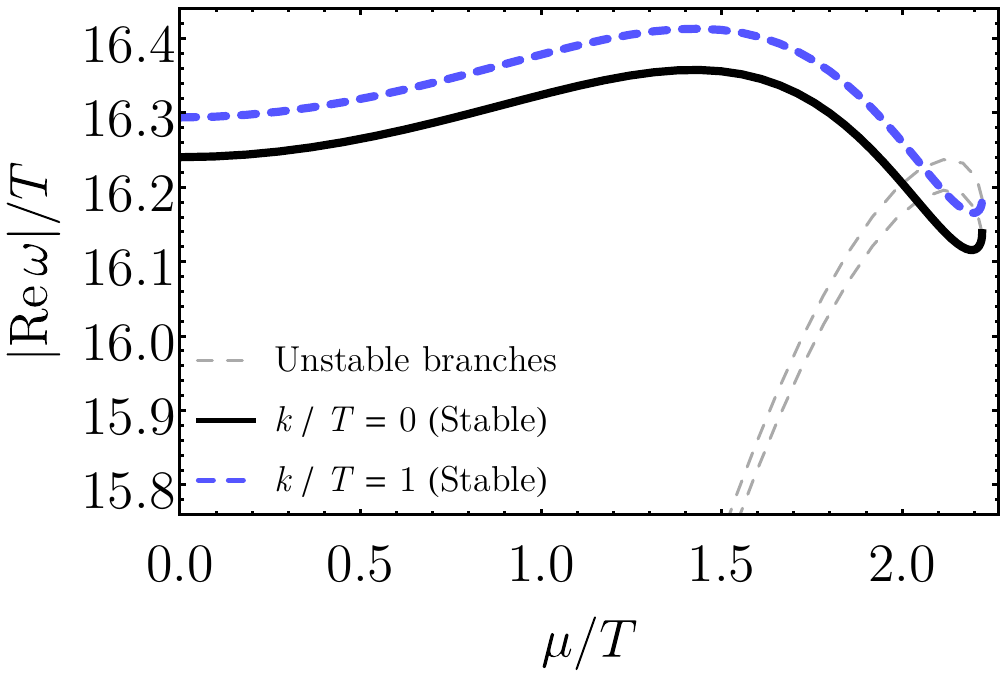}
	\caption{2nd QNM}
\end{subfigure}
\begin{subfigure}{0.49\textwidth}	
	\includegraphics[width=\textwidth]{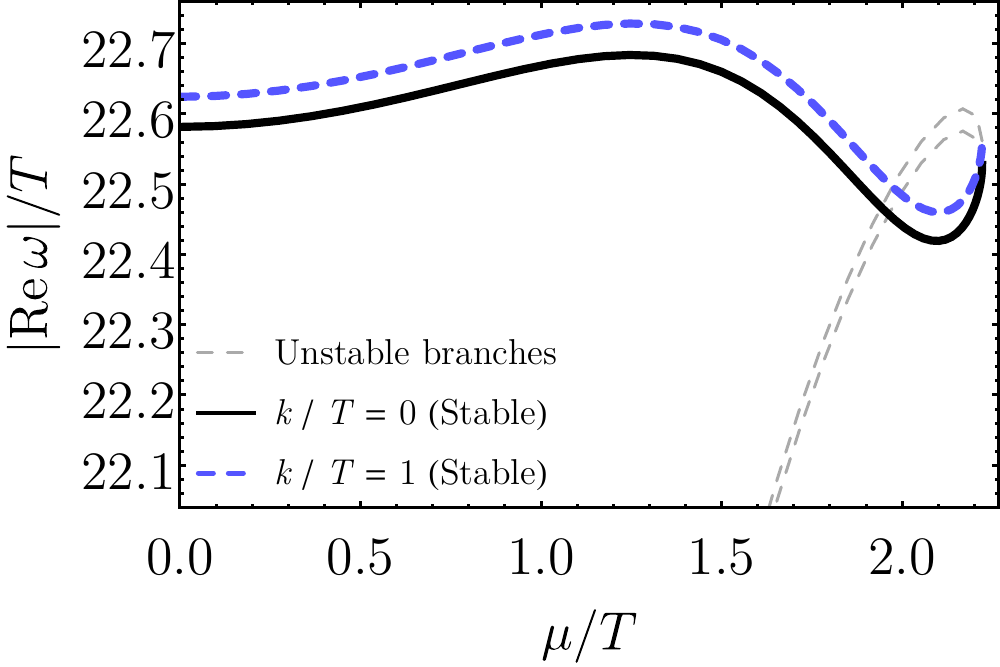}
	\caption{3rd QNM}
\end{subfigure}
\begin{subfigure}{0.49\textwidth}	
	\includegraphics[width=\textwidth]{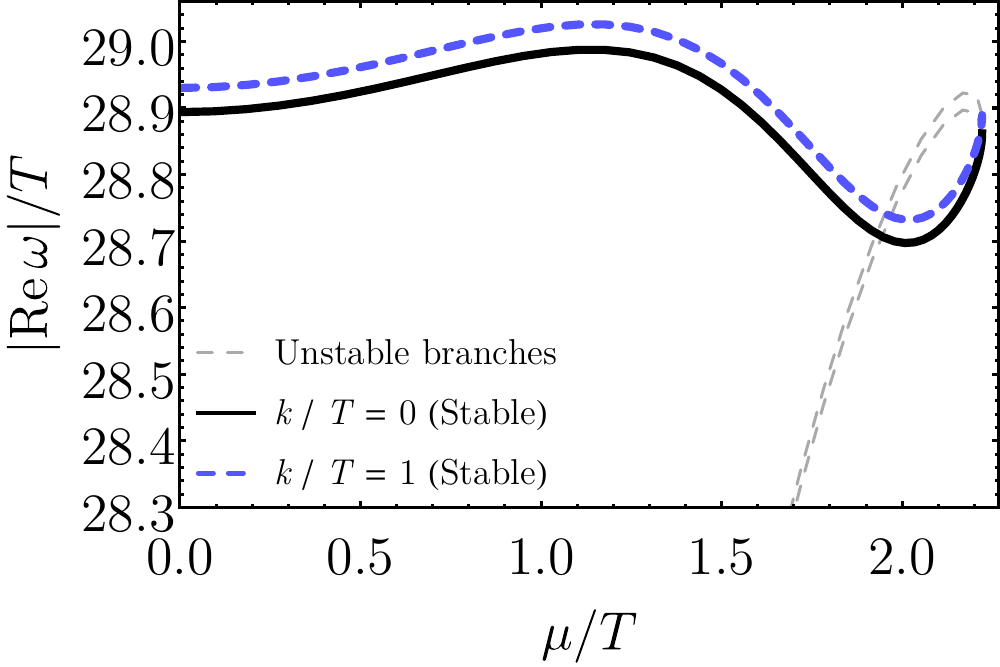}
	\caption{4th QNM}
\end{subfigure}
\caption{(Color online) Absolute value of the real part of the first 4 QNM's in the external scalar channel for $k/T = 0$ and $k/T = 1$, as a function $\mu/T$, for both stable and unstable branches.}
\label{fig:qnmre}
\end{figure}

In Figs.\ \ref{fig:qnmim} and \ref{fig:qnmre} we display the imaginary and real parts of the first 4 QNM's as functions of $\mu/T$, for both stable and unstable branches, at $k/T = 0$ and $k/T = 1$. We see that at the critical point all the QNM's develop an infinite slope. Moreover, we also note that the effects on the non-hydrodynamic modes due to finite momentum are small for $k/T\sim 1$ (especially for the imaginary part), being more pronounced for the lowest QNM's, which seems to be a general holographic property of the dispersion relation of non-hydrodynamics QNM's known as ``ultralocality'' \cite{Heller:2014wfa,Janik:2015waa}.

\begin{figure}[t]
\centering
\begin{subfigure}{0.49\textwidth}
\includegraphics[width=\textwidth]{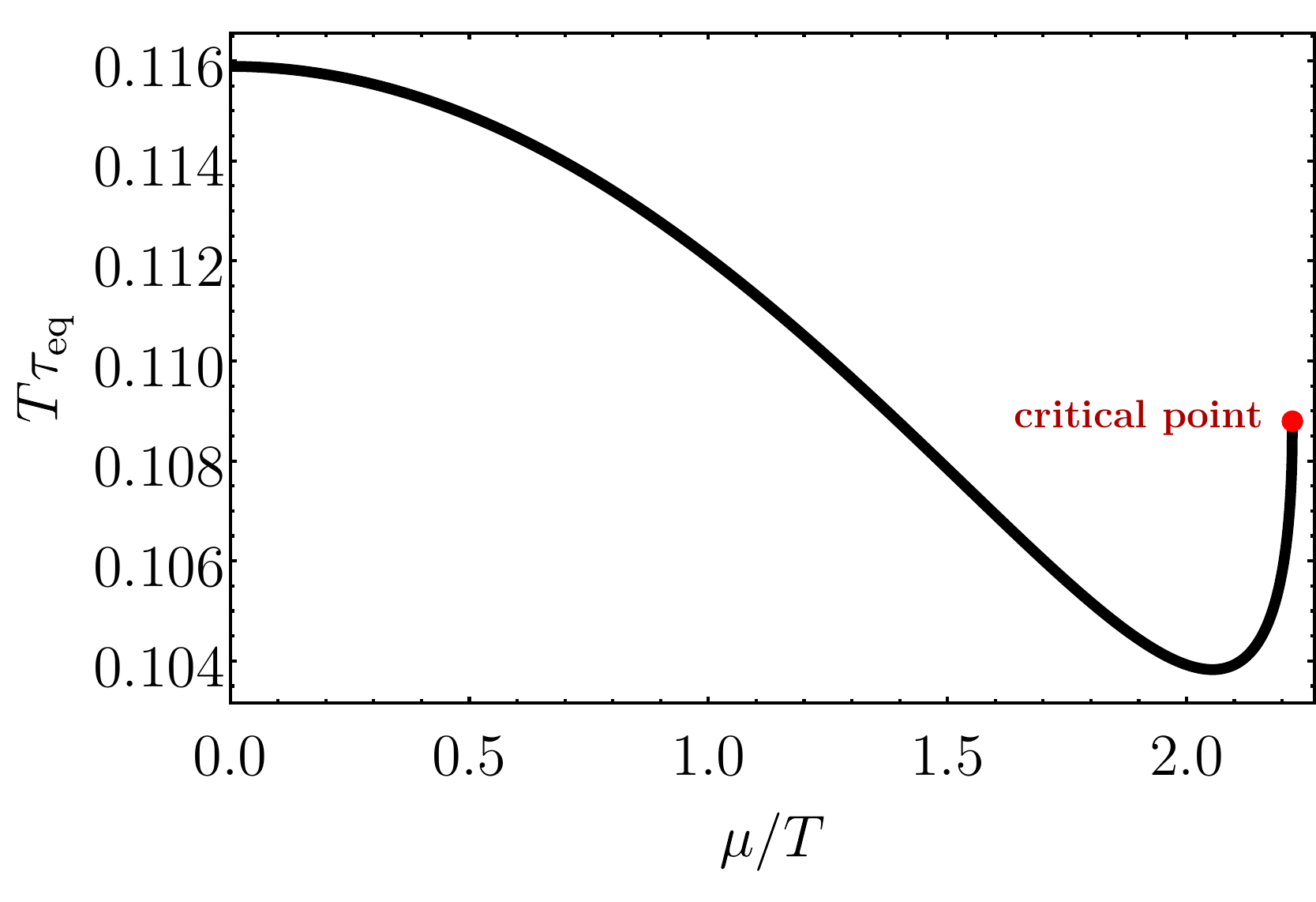}
\caption{}
\end{subfigure}
\begin{subfigure}{0.49\textwidth}
\includegraphics[width=\textwidth]{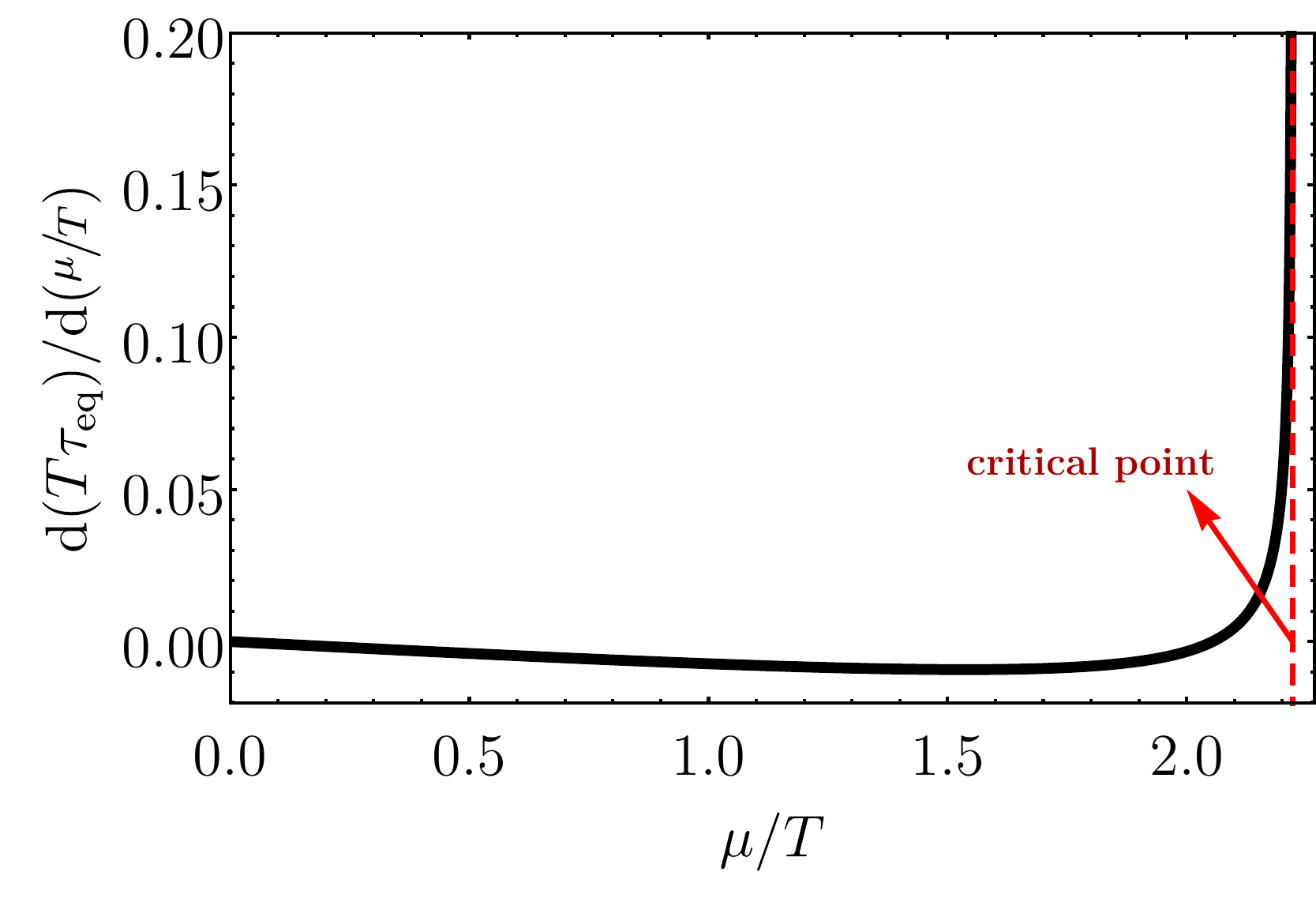}
\caption{}
\end{subfigure}
\caption{(Color online) Equilibration time (a) and its normalized derivative (b) in the external scalar channel as functions of $\mu/T$ at zero wavenumber.}
\label{fig:eqtime}
\end{figure}

Following \cite{Horowitz:1999jd}, one may define an upper bound for the equilibration time of the plasma, $\tau_\mathrm{eq}$, using the inverse of minus the imaginary part of the lowest non-hydrodynamical QNM evaluated at zero momentum. This is shown in Fig.\ \ref{fig:eqtime} (a), from which one can see that far from the critical point the equilibration time of the finite $U(1)$ R-charge density SYM plasma decreases with increasing chemical potential, again in qualitative agreement with what was previously found in \cite{Rougemont:2015wca} in the context of a phenomenologically realistic holographic model for the QGP at finite baryon chemical potential. However, for larger chemical potentials, as one approaches the critical point of the model, this behavior is modified and the equilibration time starts to increase, acquiring an infinite slope at the critical point.

We may associate a new critical exponent with the derivative of the equilibration time, $d(T\tau_\mathrm{eq})/d(\mu/T)$, since it diverges at the critical point, as shown in Fig.\ \ref{fig:eqtime} (b). Close to the critical point,
\begin{equation}
\label{eq:fit}
\dfrac{d(T\tau_\mathrm{eq})}{d\left(\mu/T\right)} \sim \left(\dfrac{\pi}{\sqrt{2}}-\dfrac{\mu}{T} \right)^{-\theta},
\end{equation}
where $\theta$ is the dynamical critical exponent which we want to calculate. By following the procedure discussed in Section \ref{sec:error}, we performed numerical fits of Eq.\ \eqref{eq:fit} to the data shown in Fig.\ \ref{fig:eqtime} (b) for different sets of intervals in $\mu/T$ increasingly closer to the critical point at $\mu/T = \pi/\sqrt{2}$. The final result gives a critical exponent compatible with $\theta = 1/2$. 

More generally, one may consider different characteristic equilibration times of the medium associated with the different non-hydrodynamic QNM's, where equilibration times associated with higher order modes should be understood as estimates for how fast the system relaxes to equilibrium depending on how rapidly varying are the perturbations to which it is subjected. Operationally, this amounts for computing the inverse of minus the imaginary part of the different non-hydrodynamic QNM's. By doing so, we obtain the same dynamical critical exponent $\theta = 1/2$ associated with all the different characteristic equilibration times of the plasma in the external scalar channel at zero wavenumber.

\section{QNM's in the vector diffusion channel}
\label{sec:vector}

\subsection{Equation of motion}

\hspace{5 mm} In this section we compute the QNM's of the vector diffusion channel in the long wavelength limit. Differently from what was done in the last section where we considered an external scalar perturbation on top of the 1RCBH background, now we need to consider fluctuations of the Maxwell field $A_{\mu}$ which is already nonzero in the background and, therefore, we also need to consider disturbances in the background metric $g_{\mu\nu}$ and dilaton field $\phi$. At zero spatial momentum the different channels for these fluctuations, at the linearized level, are classified by different representations of the $SO(3)$ rotation group \cite{DeWolfe:2011ts}.\footnote{At nonzero $k$ such classification is no longer valid and the corresponding fluctuations are organized in a more complicated way under a smaller $SO(2)$ symmetry group. We are not going to pursue the investigation of this more involved case in the present work.} By taking the fluctuation of the gauge field along the $z$ direction one finds that at the linearized level it only mixes with the fluctuation of $g^z_t$. Taking now the long wavelength limit, i.e. $k=0$, we write down for the Fourier modes of these fluctuations,
\begin{align}
\delta A_z = a(r) e^{-i \omega t}, \,\,\, \delta g^z_t = s(r) e^{-i \omega t}.
\end{align}
Then, the linearized Maxwell's equations, $\nabla_\mu\left(f(\phi)F^{\mu\nu}\right) = 0$, expressed in Eddington-Finkelstein coordinates read,
\begin{align}
- \Phi'' + \left( -2 A' + B' - \frac{f'(\phi) \phi'} {f(\phi)} \right) \Phi' & = 0, \label{eq:eom1} \\
a'' + \left( 2 A' - B' + \frac{h'}{h} -2i \omega \frac{\e{A-B}}{h} + \frac{f'(\phi) \phi'} {f(\phi)} \right) a' - i \omega \e{B-A} \frac{f(\phi) A' + f'(\phi) \phi'}{f(\phi) h} a + & \nonumber \\
+  \frac{\Phi'}{2h'} s'  +\left[\frac{f'(\phi) \phi' \Phi' + f(\phi) \big( (2A' - B') \Phi' + \Phi'' \big)}{2f(\phi) h} \right] s & = 0 \label{eq:eom2},
\end{align}
where the first equation is the equation of motion for the background Maxwell field, $\Phi(r)$, while the second equation is the equation of motion for the Maxwell perturbation, $a(r)$. We may decouple the perturbations $a(r)$ and $s(r)$ by using Einstein's equations for the metric field,
\begin{equation}
\label{eq:ein}
R_{\mu \nu} - \frac{g_{\mu \nu}}{3} \left(V(\phi) - \frac{f(\phi)}{4} F_{\alpha\beta}^2 \right)-\frac{f(\phi)}{2}F_{\mu\alpha}F_{\nu}^{\ \alpha} - \frac{1}{2} \partial_{\mu} \phi \partial_{\nu} \phi = 0.
\end{equation}
By taking the $vz$-component minus the $rz$-component of the above equation of motion, one obtains the constraint,
\begin{equation}
\label{eq:eom3}
s' = - f(\phi) \Phi' \e{-2A} a.
\end{equation}
By using the zeroth order Eq.\ \eqref{eq:eom1} to eliminate the $s(r)$ term from Eq.\ \eqref{eq:eom2} and substituting \eqref{eq:eom3} into \eqref{eq:eom2}, one obtains a decoupled equation of motion for the radial profile of the vector field perturbation, which is associated with the diffusion of the $U(1)$ R-charge \cite{Kovtun:2005ev,DeWolfe:2011ts},
\begin{align}
\label{eq:eomfinal}
a'' + \left[ 2 A' - B' + \frac{h'}{h} + \frac{f'(\phi) \phi'} {f} - 2i \omega \frac{\e{B-A}}{h} \right] a' + & \nonumber \\  - \frac{\e{-2A}}{h} \left[ i \omega \e{A+B} \left(  A' + \phi' \frac{f'(\phi)}{f(\phi)} \right) + f(\phi) \Phi'^2 \right] a & = 0.
\end{align}

Once again we apply the radial coordinate transformation $r \to r_H/u$, which yields,
\begin{align}
a'' & + \left(-\frac{10 u^6 (y-3) (y-1)-3 u^4 (5 (y-2) y+1)+4 u^2 \left(y^2-1\right)+(y+1)^2}{u \left(u^2-1\right) \left(u^2(y-3)-y-1\right) \left(-2 u^2 (y-1)+y+1\right)^2}+ \right. \nonumber \\ & \left.  -\frac{4 i (\omega/T) \sqrt{(3-y) \left(-2 u^2 (y-1)+y+1\right)}}{\pi  \left(u^2-1\right) (y-3) \left(u^2 (y-3)-y-1\right)}\right)a' +  \nonumber \\ & + \left(-\frac{8 u^4 (y-3) (y-1) (y+1)}{\left(u^2-1\right) \left(u^2 (y-3)-y-1\right) \left(-2 u^2 (y-1)+y+1\right)^2} + \right.  \nonumber \\ & \left. -\frac{2 i (\omega/T) \left(4 u^2 (y-1)+y+1\right)}{\pi  \left(u^2-1\right) u \left(u^2 (y-3)-y-1\right) \sqrt{(y-3) \left(2 u^2 (y-1)-y-1\right)}}\right) a = 0.
\end{align}
For the vector perturbation the normalizable mode at the boundary corresponds to set $a(u) = u^2 F(u)$, with $F(0)\neq 0$, from which one finally obtains, 
\begin{align}
F'' & + \left[-\frac{18 u^6 (y-3) (y-1)-7 u^4 (5 (y-2) y+1)+20 u^2 \left(y^2-1\right)-3 (y+1)^2}{\left(u^2 (y-3)-y-1\right) \left(-2 u^2 (y-1)+y+1\right)} + \nonumber  \right. \\ & \left. -\frac{4 i u (\omega/T) \sqrt{(3-y) \left(-2 u^2 (y-1)+y+1\right)}}{\pi  (y-3) \left(u^2 (y-3)-y-1\right)}\right] \frac{1}{u(u^2-1)} F' + \nonumber \\ & +  \frac{1}{u(u^2-1)} \left[\frac{8 u \left(6 u^6 (y-3) (y-1)^2-2 u^4 (y (7 (y-3) y+9)+5)\right)}{\left(u^2 (y-3)-y-1\right)  \left(-2 u^2 (y-1)+y+1\right)^2} \right. \nonumber \\ & \left. + \frac{8u \left(u^2 (y+1) (3 y-5) (3 y-1)-2 (y-1) (y+1)^2 \right)}{\left(u^2 (y-3)-y-1\right)  \left(-2 u^2 (y-1)+y+1\right)^2} \right. \nonumber \\ & \left. -\frac{6 i (\omega/T) \left(4 u^2 (y-1)-y-1\right)}{\pi  \left(u^2 (y-3)-y-1\right) \sqrt{(y-3) \left(2 u^2 (y-1)-y-1\right)}}\right] F = 0.
\end{align}

\subsection{QNM spectra and equilibration time}
\label{sec:QNMdiff}

\begin{figure}[t]
\centering
\includegraphics[width=0.7\textwidth]{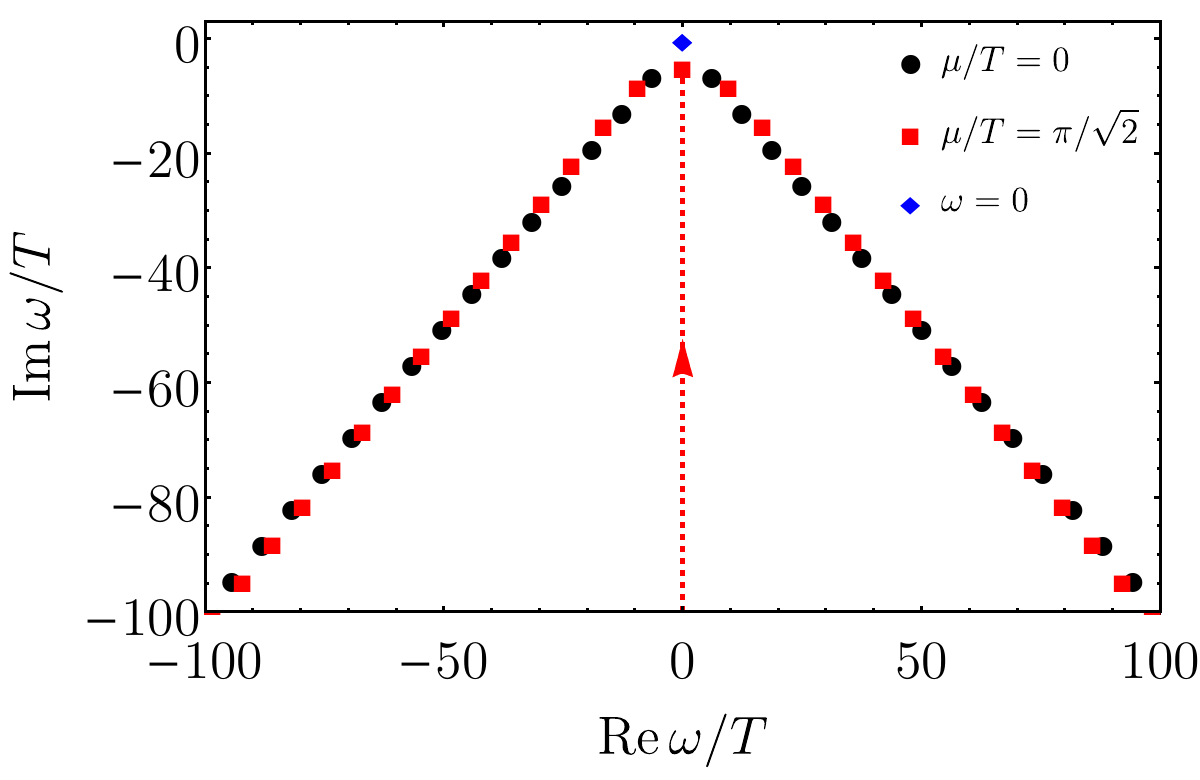}
\caption{(Color online) QNM spectra of the first 30 symmetric poles in the vector diffusion channel for $\mu/T = 0$ (black circles) and $\mu/T = \pi/\sqrt 2$ (red squares) at $k/T=0$. The hydrodynamical diffusive pole is depicted by the blue diamond. Note also the emergence of a new purely imaginary, non-hydrodynamical mode which comes from $-i\infty$ at $\mu/T=0$ and remains at a finite distance from the origin at the critical point $\mu/T = \pi/\sqrt 2$.}
\label{fig:qnmtypcond}
\end{figure}

\begin{figure}[t]
\centering
\begin{subfigure}{0.49\textwidth}	
	\includegraphics[width=\textwidth]{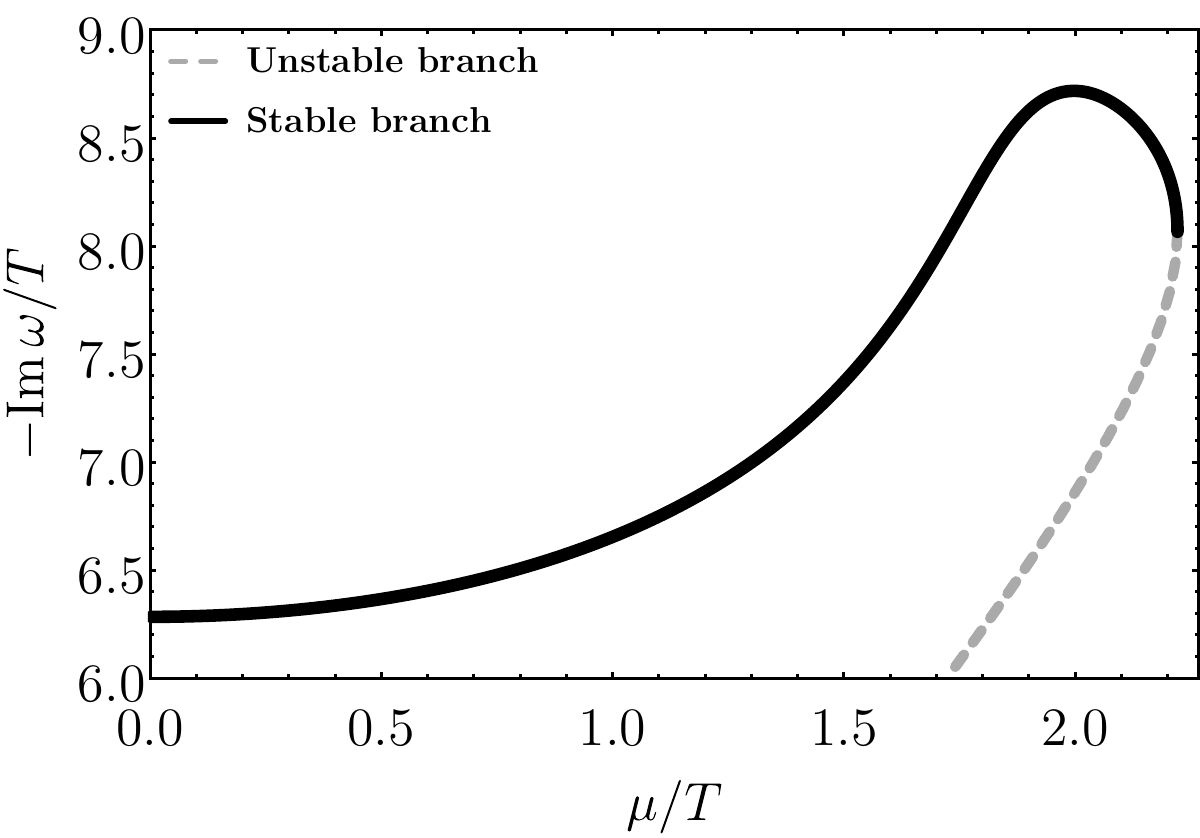}
	\caption{1st QNM}
\end{subfigure}
\begin{subfigure}{0.49\textwidth}	
	\includegraphics[width=\textwidth]{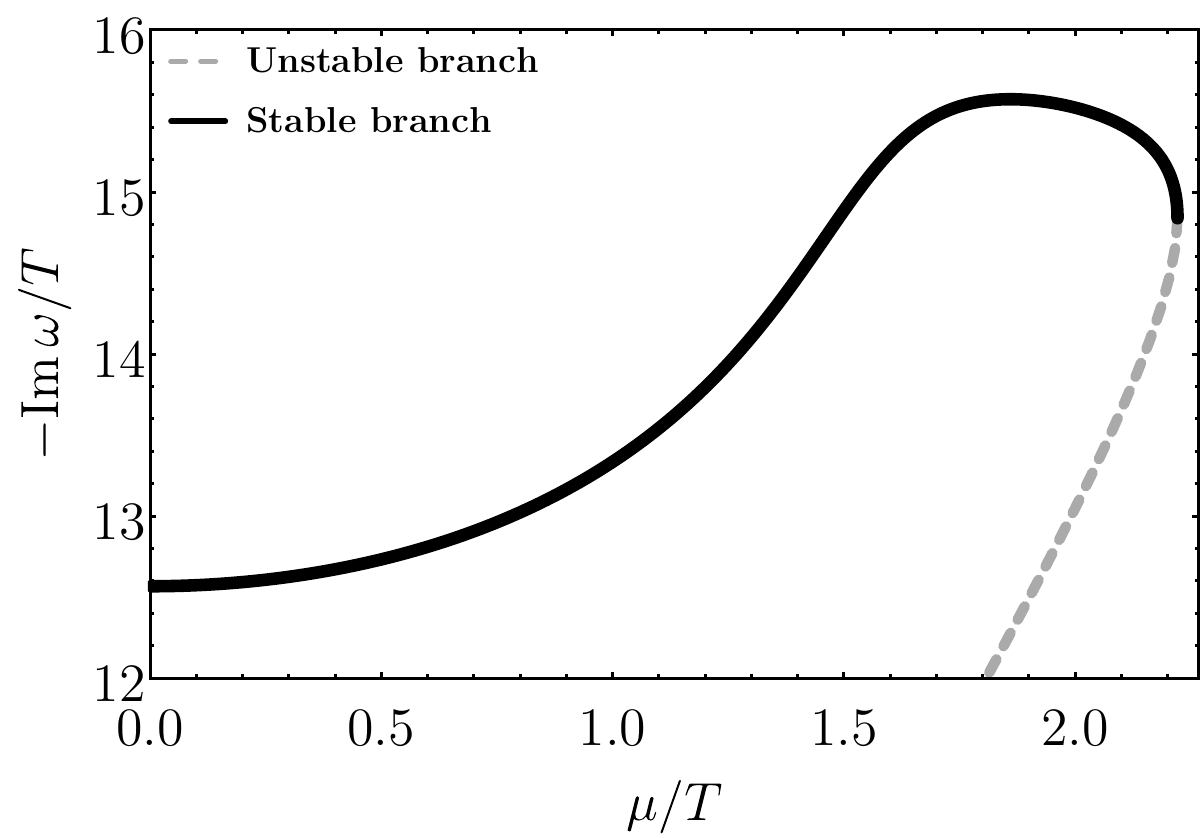}
	\caption{2nd QNM}
\end{subfigure}
\begin{subfigure}{0.49\textwidth}	
	\includegraphics[width=\textwidth]{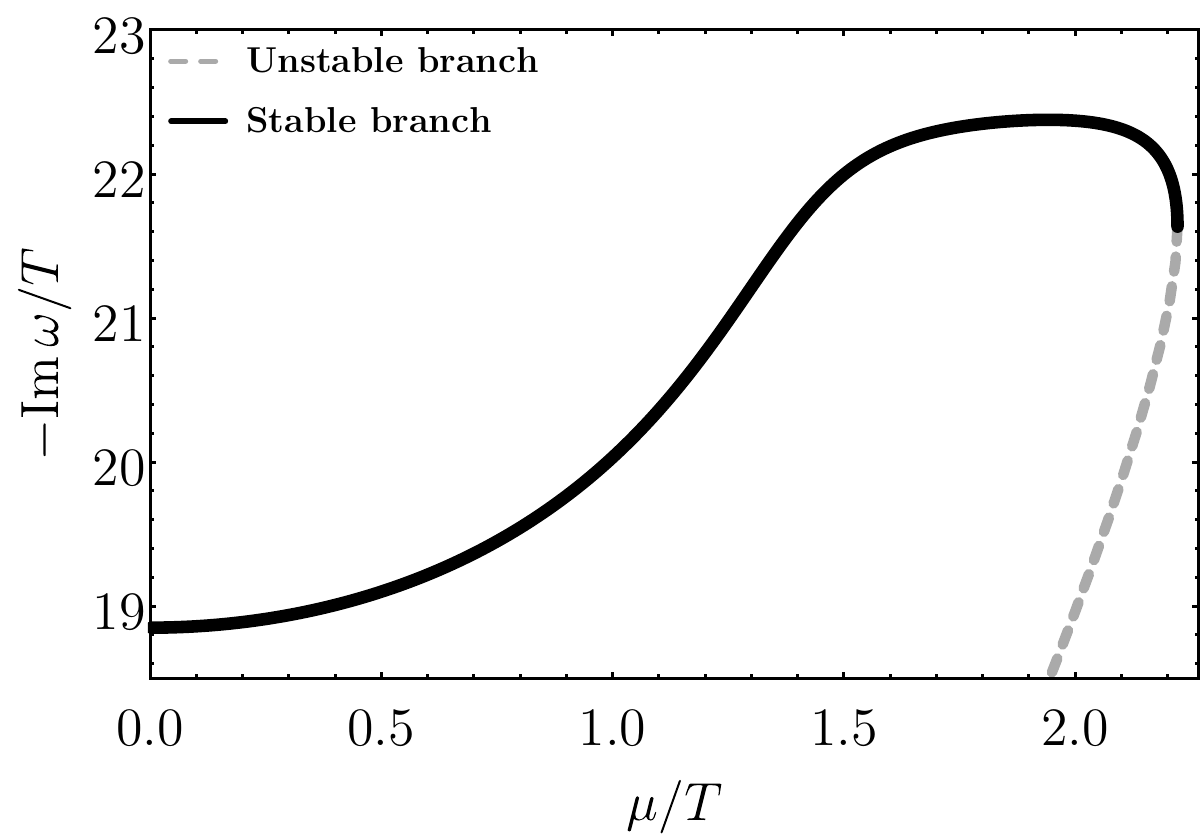}
	\caption{3rd QNM}
\end{subfigure}
\begin{subfigure}{0.49\textwidth}	
	\includegraphics[width=\textwidth]{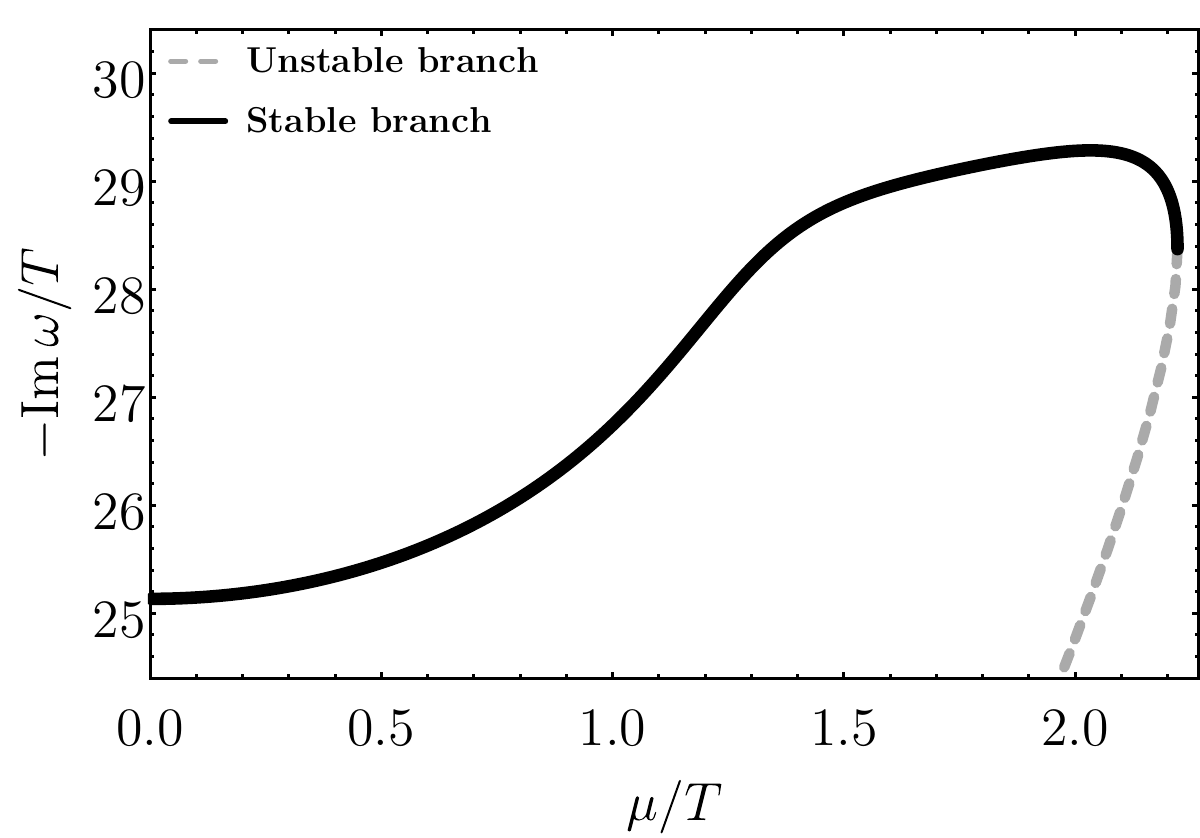}
	\caption{4th QNM}
\end{subfigure}
\caption{(Color online) Imaginary part of the first 4 QNM's in the vector diffusion channel for $k/T = 0$, as a function of $\mu/T$, for both stable and unstable branches.}
\label{fig:qnmimcond}
\end{figure}

\begin{figure}[t]
\centering
\begin{subfigure}{0.49\textwidth}	
	\includegraphics[width=\textwidth]{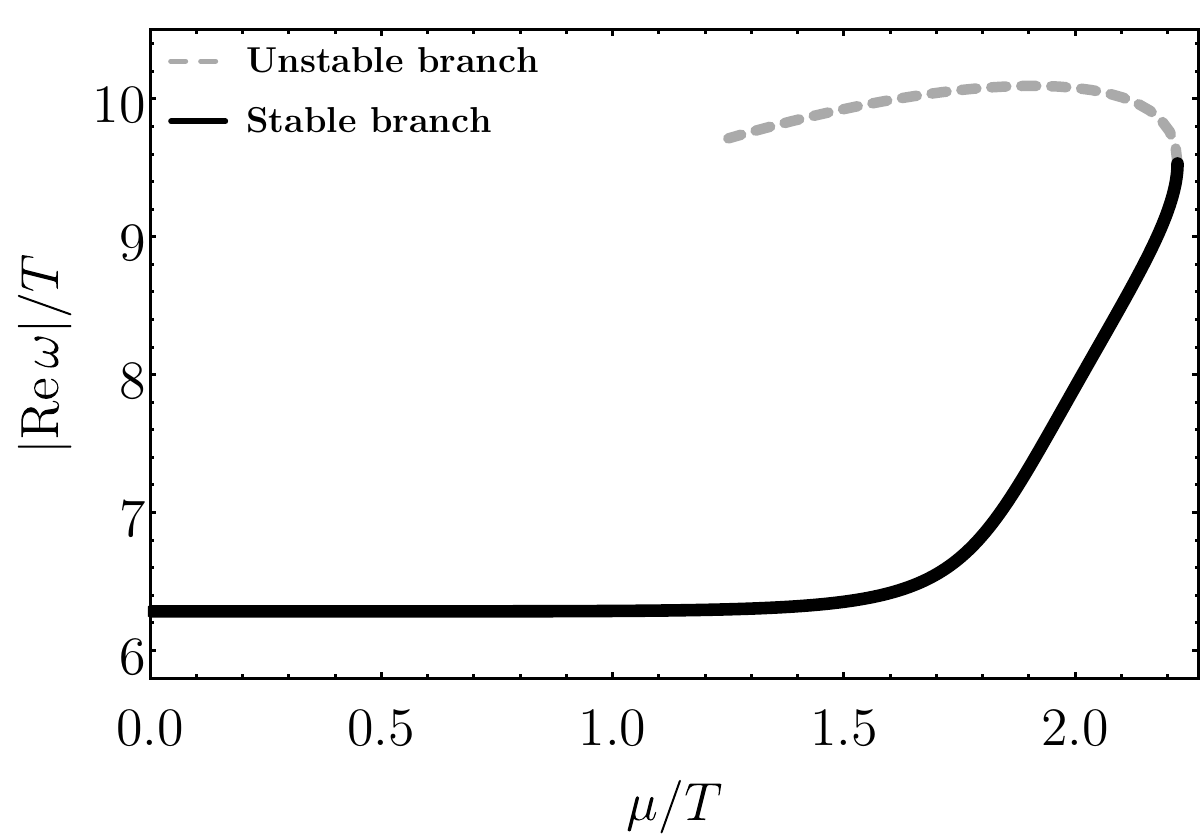}
	\caption{1st QNM}
\end{subfigure}
\begin{subfigure}{0.49\textwidth}	
	\includegraphics[width=\textwidth]{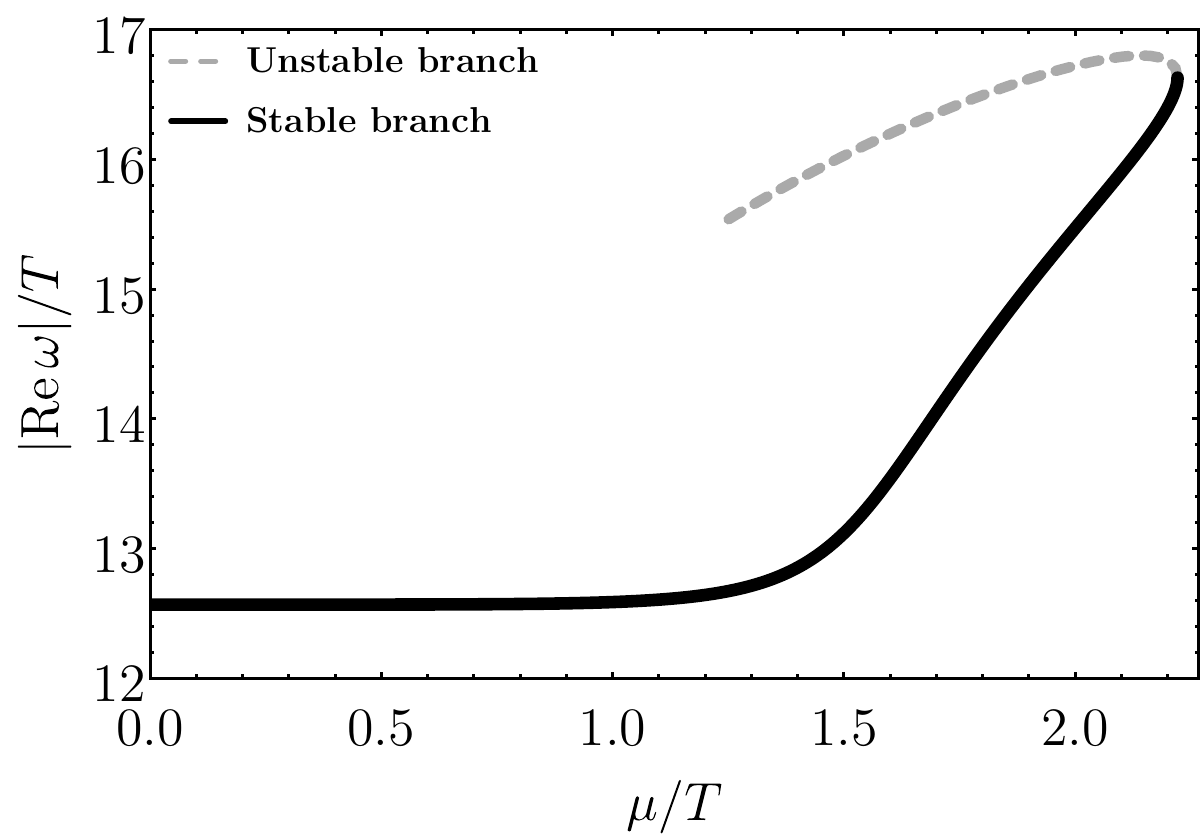}
	\caption{2nd QNM}
\end{subfigure}
\begin{subfigure}{0.49\textwidth}	
	\includegraphics[width=\textwidth]{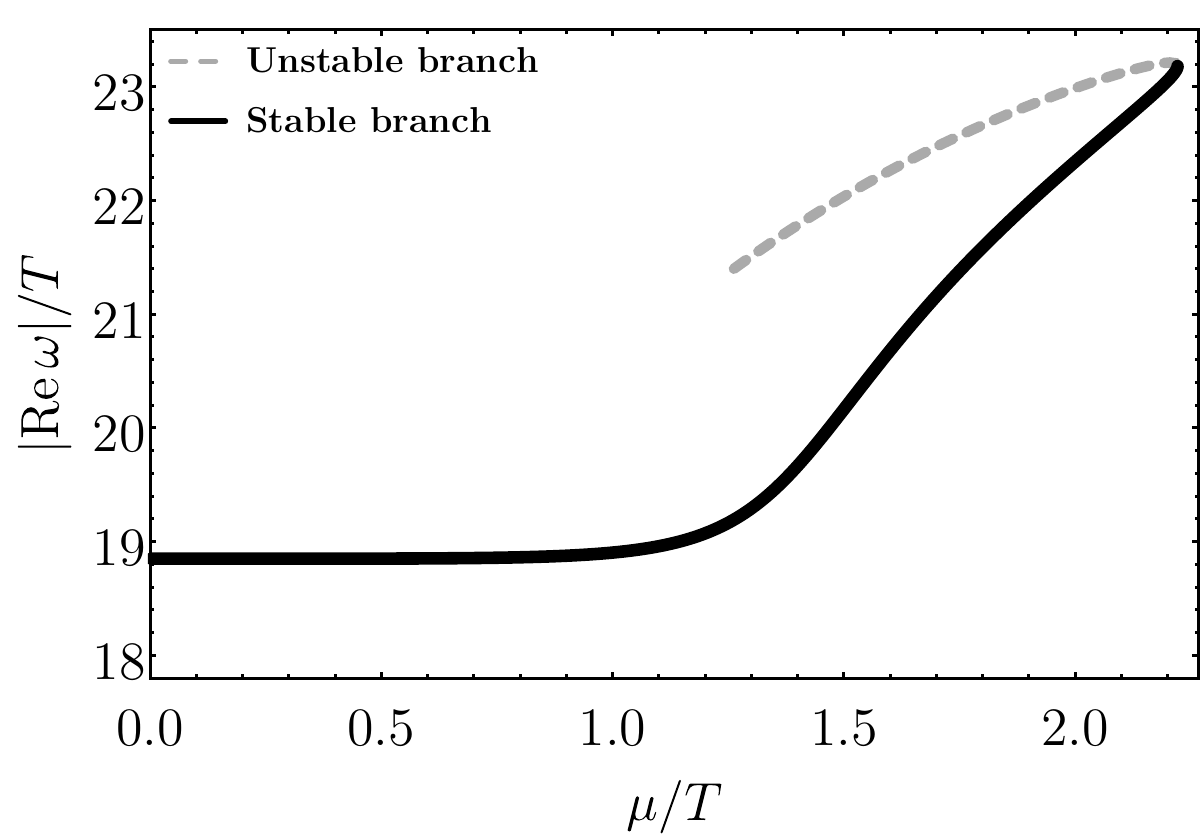}
	\caption{3rd QNM}
\end{subfigure}
\begin{subfigure}{0.49\textwidth}	
	\includegraphics[width=\textwidth]{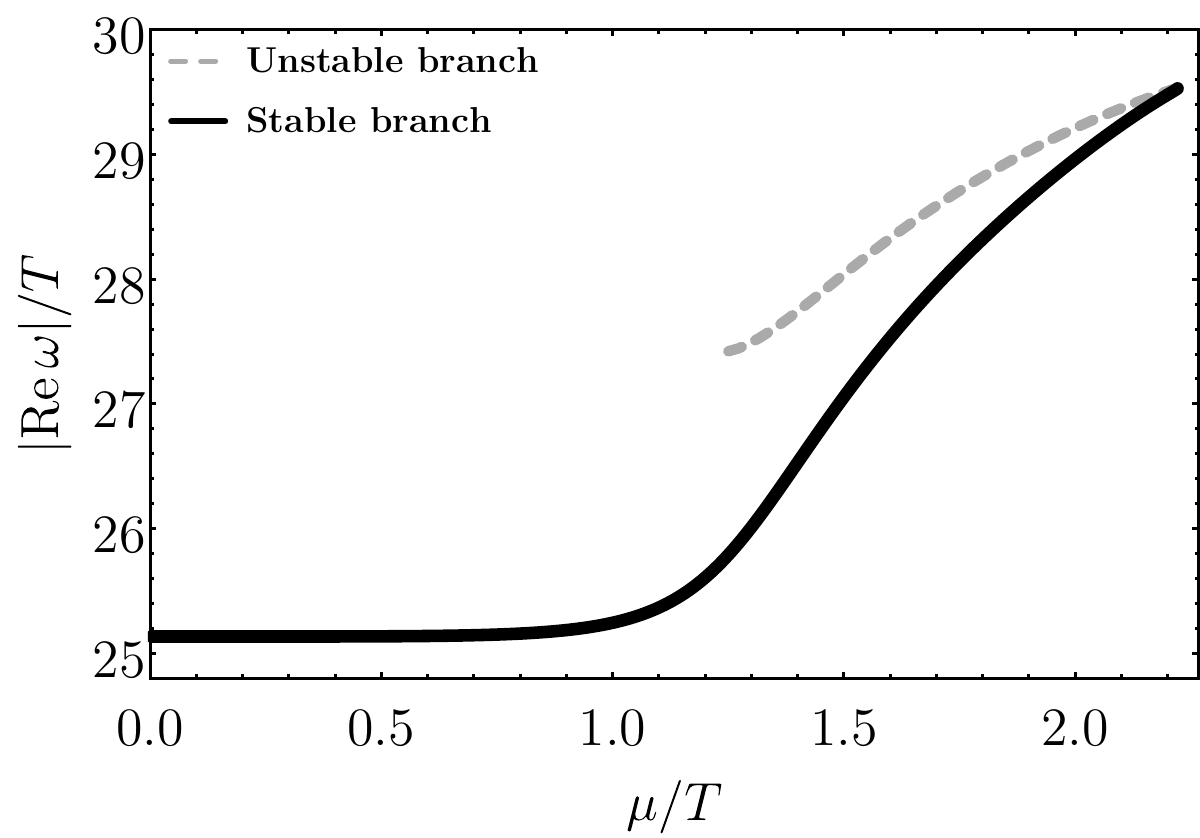}
	\caption{4th QNM}
\end{subfigure}
\caption{(Color online) Absolute value of the real part of the first 4 QNM's in the vector diffusion channel for $k/T = 0$, as a function $\mu/T$, for both stable and unstable branches.}
\label{fig:qnmrecond}
\end{figure}

\hspace{5 mm} With the QNM eigenvalue problem completely specified as discussed above, we can now apply the pseudospectral method (see Appendix \ref{sec:numerics}) to numerically solve it. In Fig.\ \ref{fig:qnmtypcond} we display the QNM spectra for the first 30 symmetric poles in the vector diffusion channel in the limiting cases of $\mu/T = 0$ (\AdS{5}-Schwarzschild) and $\mu/T = \pi/\sqrt 2$ (critical point). In Figs.\ \ref{fig:qnmimcond} and \ref{fig:qnmrecond} we show the imaginary and real parts of the first 4 complex QNM's as functions of $\mu/T$, for both stable and unstable branches. We see that also in the vector diffusion channel both the real and imaginary parts of the QNM's develop an infinite slope at the critical point.

We remind the reader that at $k=\mu=0$ the QNM spectra in the vector diffusion channel may be analytically calculated \cite{Kovtun:2005ev},
\begin{equation}
\frac{\omega}{T}=2\pi n(1-i),\,\,n\in\mathbb{N}\,\,\,\textrm{at}\,\,\,\mu=0.
\end{equation}
Our numerical calculations at $\mu/T=0$ agree with this analytical result. The standard hydrodynamical mode $\omega(k/T=0)/T=0$ is depicted by the blue diamond in Fig.\ \ref{fig:qnmtypcond}. Since this is a hydrodynamical pole, it does not evolve with the chemical potential if we keep $k=0$. This mode determines the R-charge conductivity of the model and the zero frequency limit of this transport coefficient was found in \cite{DeWolfe:2011ts} to remain finite at the critical point, as expected for a type B dynamic universality class. However, it was noticed in \cite{DeWolfe:2011ts} that the derivative of this quantity near the critical point has infinite slope described by an exponent equal to $1/2$, which matches the exponent found in the previous section in the study of non-hydrodynamic modes of different nature corresponding to external scalar perturbations. 

On the other hand, the main effect of the chemical potential on the symmetric non-hydrodynamical modes is to increase the magnitude of both the imaginary and real parts of these poles. Therefore, also in the vector diffusion channel one sees that the inclusion of a chemical potential leads to additional damping for the quasinormal black hole oscillations.

A novel feature we observe in Fig.\ \ref{fig:qnmtypcond} is the emergence of a new purely imaginary, non-hydrodynamical pole at finite chemical potential, which comes from $\omega/T\to-i\infty$ at $\mu/T=0$ and lies at $\omega/T\approx -7.315i$ at the critical point $\mu/T = \pi/\sqrt 2$. For $\mu/T\gtrsim 2$, this new purely imaginary pole becomes the lowest non-hydrodynamical mode, while for lower values of the chemical potential the lowest non-hydrodynamical mode is given by any of the first two symmetric poles with respect to the imaginary axis. Therefore, this new non-hydrodynamical imaginary mode plays a crucial role in the description of the equilibration time of the system in the vector diffusion channel when the chemical potential is large, specially at criticality, when it dominates the physics of the slowest varying perturbations. The appearance of such a purely imaginary mode is an interesting feature of this model that shows that the distinction between transient phenomena at weak and strong coupling, currently understood in terms of their different pattern of non-hydrodynamic modes at zero wavenumber \cite{Denicol:2011fa} (see also \cite{Noronha:2011fi}) corresponding to fluctuations around global equilibrium, can become more complicated near a critical point.

\begin{figure}[t]
\centering
\begin{subfigure}{0.49\textwidth}
\includegraphics[width=\textwidth]{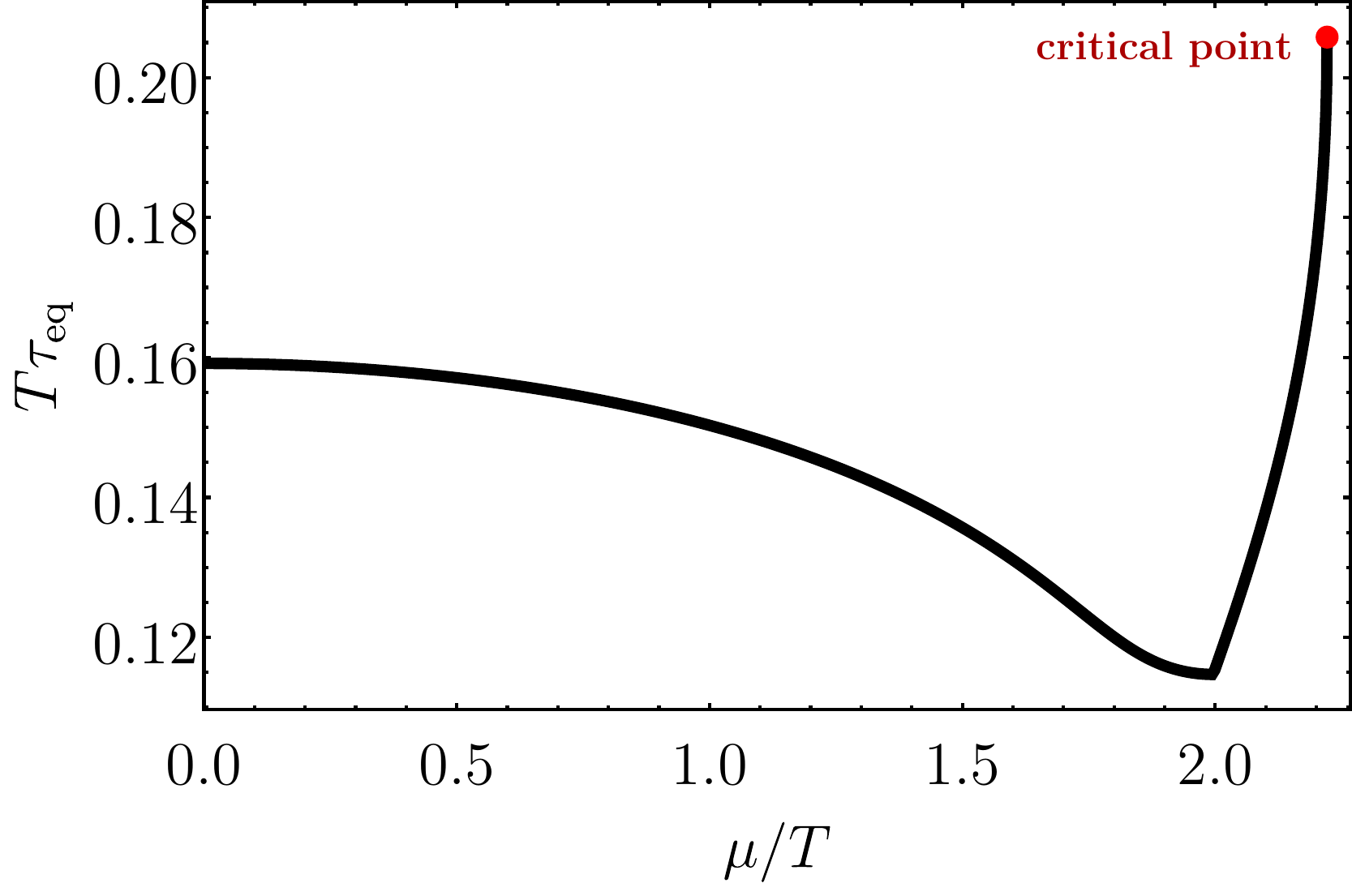}
\caption{}
\end{subfigure}
\begin{subfigure}{0.49\textwidth}
\includegraphics[width=\textwidth]{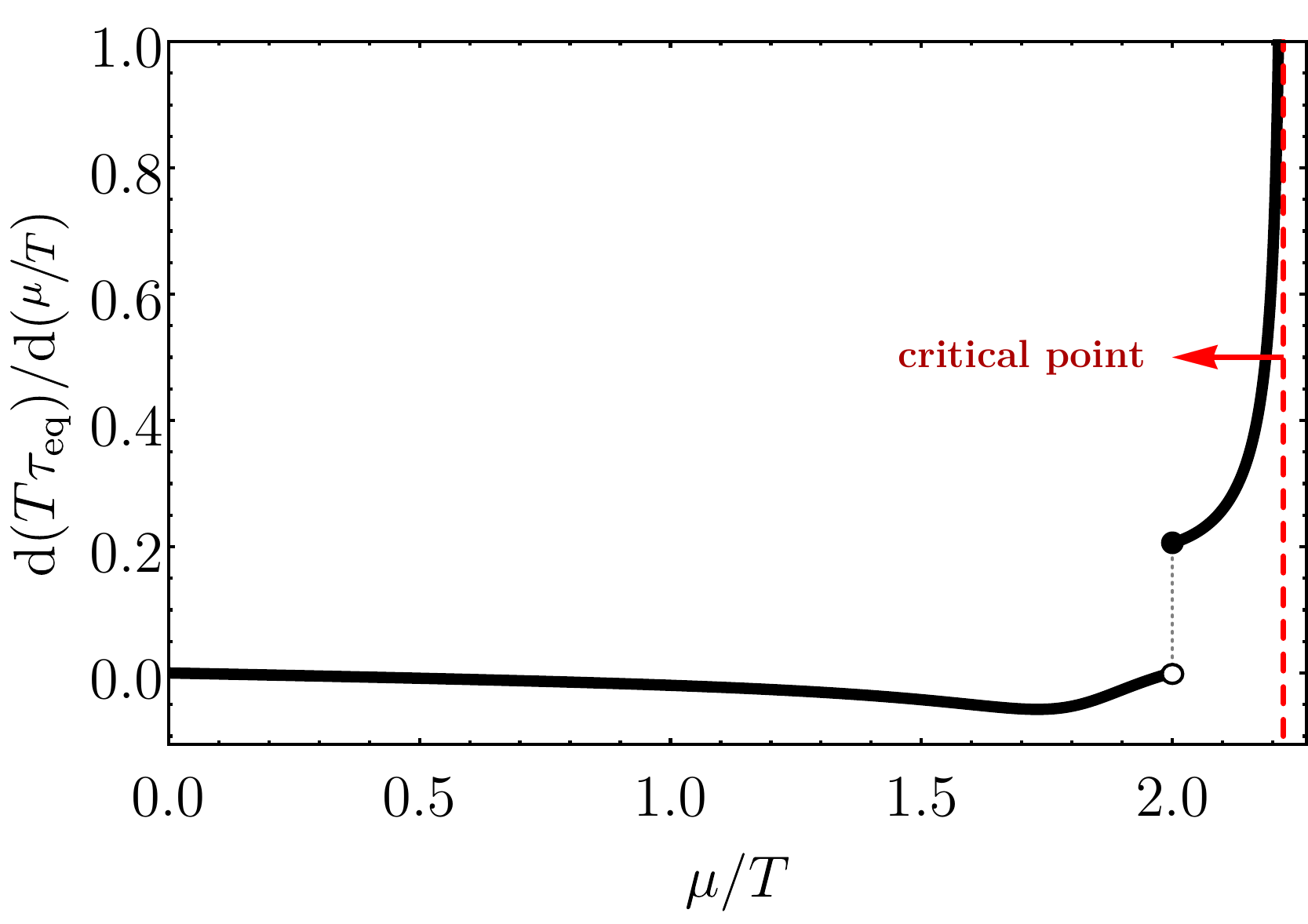}
\caption{}
\end{subfigure}
\caption{(Color online) Equilibration time (a) and its normalized derivative (b) in the vector diffusion channel as functions of $\mu/T$.}
\label{fig:eqtimecond}
\end{figure}

We define the upper bound for the equilibration time of the system in the vector diffusion channel as before by taking the inverse of minus the imaginary part of the lowest non-hydrodynamical QNM. The result is shown in Fig.\ \ref{fig:eqtimecond} (a). The kink observed in the equilibration time at $\mu/T \approx 2$ is due to the shift from the regime dominated by the first symmetric poles to the regime dominated by the new purely imaginary mode. This also causes a discontinuity in the derivative of the equilibration time, as seen in Fig.\ \ref{fig:eqtimecond} (b). As before, one can calculate the critical exponent associated with this derivative at the critical point and the result is once again compatible with $\theta=1/2$. This shows that in this model both the hydrodynamic and the non-hydrodynamic modes in this vector diffusion channel have the same critical exponents. 

\section{Final Remarks}
\label{sec:remarks}

\hspace{5 mm} In the present work we investigated the behavior of the QNM's for the external scalar and vector diffusion channels of the 1RCBH model, which is a top-down gauge/gravity construction dual to a SYM plasma at nonzero $U(1)$ R-charge density. We analyzed the behavior of the QNM's through the phase diagram of the model, which displays a critical point at a second order phase transition. We found that, except close to the critical point, by increasing the chemical potential one generally increases the damping of the quasinormal black hole oscillations, which leads to a reduction of the characteristic equilibration times of the dual plasma. However, as one approaches the critical point these equilibration times are enhanced and they acquire an infinite slope at the criticality. We found that the derivatives of all the characteristic equilibration times of the medium, obtained from the non-hydrodynamic QNM's at zero wavenumber, share the same critical exponent $\theta=1/2$. Previously, the same value was also found for the critical exponent associated to the derivative of the DC conductivity extracted from the R-charge diffusive hydrodynamic mode in this model \cite{DeWolfe:2011ts}. We also found a purely imaginary, non-hydrodynamical mode in the vector diffusion channel at nonzero chemical potential which dictates the critical behavior of the equilibration time in this channel.

The observation that the quasinormal black hole oscillations away from the critical region are additionally damped by a nonzero chemical potential, obtained here for a top-down conformal construction dual to a SYM plasma at finite R-charge density, is consistent with the behavior found previously in \cite{Rougemont:2015wca} for a rather different holographic construction, which involves a bottom-up model with black brane solutions that are engineered to describe the realistic non-conformal physics of the QGP both at zero baryon density \cite{Finazzo:2014cna} and also at finite density \cite{Rougemont:2015ona}. This may indicate that this additional damping in the quasinormal black hole oscillations due to a nonzero chemical potential, and the consequent attenuation of the equilibration time of the dual plasma away from criticality, may be a general holographic property of strongly correlated quantum fluids.

Regarding the purely imaginary, non-hydrodynamical mode found in the diffusion channel at finite density, the common link between the 1RCBH model studied here and the Einstein-Maxwell model investigated in Ref. \cite{Janiszewski:2015ura} (where this mode was also found, although with no critical behavior, due to the lack of a phase transition in the model of Ref. \cite{Janiszewski:2015ura}) is the presence of bulk electromagnetic fields. Somehow, the Maxwell field changes qualitatively the dynamical response of the system to perturbations. From the point of view of the gravitational dynamics in five dimensions, what we see is that this purely imaginary non-hydrodynamic mode appears for electromagnetically charged asymptotically AdS black holes and that, when the charge of the black hole is large enough, this new mode dominates the dynamics of relaxation towards equilibrium when the black hole is disturbed. From the point of view of the four dimensional dual quantum field theory at the boundary of the asymptotically AdS spacetime background, this new mode dominates the dynamics responsible for the characteristic equilibration time of the plasma at large enough densities in the diffusion channel. Since bulk electric fields map to boundary states at finite density, this may be a general feature of holographic models at finite density. Therefore, one could investigate if the behavior of non-hydrodynamic modes in other holographic models that display critical phenomena possess similar properties to those found in the present study, i.e., the corresponding equilibration times have infinite slope characterized by a single critical exponent $\theta$. In particular, we intend to investigate in the near future these features in two bottom-up constructions of phenomenological relevance for the physics of the QGP: the EMD model at finite baryon chemical potential \cite{Rougemont:2015wca,Rougemont:2015ona} and the anisotropic EMD model at finite magnetic field from Refs.\ \cite{Rougemont:2015oea,Finazzo:2016mhm}. Such studies may be relevant to understand how the presence of a critical endpoint in the QCD phase diagram may lead to new observables associated with, for instance, baryon transport in the baryon rich quark-gluon plasma produced in heavy ion collisions within the beam energy scan program at RHIC.

\acknowledgments

S.~I.~Finazzo was supported by the S\~{a}o Paulo Research Foundation (FAPESP) and Coordena\c{c}\~ao de Aperfei\c{c}oamento de Pessoal de N\'{i}vel Superior (CAPES) under FAPESP grant number 2015/00240-7. R.~Rougemont acknowledges financial support by FAPESP under FAPESP grant number 2013/04036-0. M.~Zaniboni acknowledges financial support by CAPES. R.~Critelli acknowledges financial support by FAPESP under FAPESP grant number 2016/09263-2. J.~Noronha acknowledges financial support by FAPESP and Conselho Nacional de Desenvolvimento Cient\'{i}fico e Tecnol\'{o}gico (CNPq).

\appendix

\section{Spectral function for the external scalar channel}
\label{sec:spectral}

\hspace{5 mm} In this Appendix we study the spectral function associated with the external scalar fluctuation in the bulk. The motivation for this pursuit comes from the fact that QNM's are associated with poles of the retarded Green's function, whose imaginary part defines the spectral function. Therefore, one should expect that the critical behavior found for the QNM's leave somehow an imprint in the spectral function. Here we investigate this issue by considering the spectral function of the external scalar field fluctuation. Also, we note that due to the universal character of the $\delta g_x^y$ fluctuation of the metric \cite{Kovtun:2004de}, the same calculation also gives the shear viscosity spectral function. 

For the sake of completeness, let us begin by briefly reviewing the holographic computation of the spectral function based on the real time holographic prescription \cite{Son:2002sd} recast using the holographic membrane paradigm \cite{Iqbal:2008by}. From linear response theory, the expectation value of the QFT operator $\hat{O}$ dual to the scalar perturbation is associated with the retarded correlator according to (see \cite{Ramallo:2013bua}),
\begin{equation}
\langle \hat{O} (\omega, \vec{k}) \rangle = - \mathcal{G}^R_{\hat{O}\hat{O}} (\omega, \vec{k}) J(\omega, \vec{k}),
\end{equation}
where, as before, $J(\omega, \vec{k})$ denotes the leading mode of the scalar fluctuation at the boundary, which sources the QFT operator $\hat{O}$, while the expectation value is associated with the subleading mode, being given by,
\begin{equation}
\langle \hat{O} (\omega, \vec{k}) \rangle = \lim\limits_{r\to\infty}\Pi (r, \omega, \vec{k}) = \lim\limits_{r\to\infty} \frac{\delta S}{\delta\tilde{\varphi}'},
\end{equation}
where $\Pi$ is the radial canonical momentum conjugate to the scalar perturbation $\tilde{\varphi}$. Once more, we impose that the scalar perturbation satisfies an in-falling wave condition at the horizon, which gives the retarded propagator.

Our goal here is to compute the spectral function defined by,
\begin{align}\label{eq:spectral}
F_s\equiv-\Im \mathcal{G}^R_{\hat{O}\hat{O}}=\lim\limits_{r\to\infty}\Im\dfrac{\Pi}{\tilde{\varphi}}.
\end{align}
To do so, we have to first solve the equation of motion that follows from the action \eqref{eq:scalaraction} in the usual coordinates of \eqref{eq:line element},
\begin{equation}
\label{eq:eomspectral1}
\tilde{\varphi} '' + \left(4 A'-B'+\frac{h'}{h}\right) \tilde{\varphi}' +\dfrac{\e{2(B-A)}}{h^2}\left(\omega^2-k^2h\right) \tilde{\varphi} = 0,
\end{equation}
with an in-falling horizon condition at $r = r_H$ and $\lim\limits_{r\to\infty}\tilde{\varphi}(r, \omega, \vec{k}) = J(\omega,\vec{k})$.

\begin{figure}[t]
\centering
\includegraphics[width=0.7\textwidth]{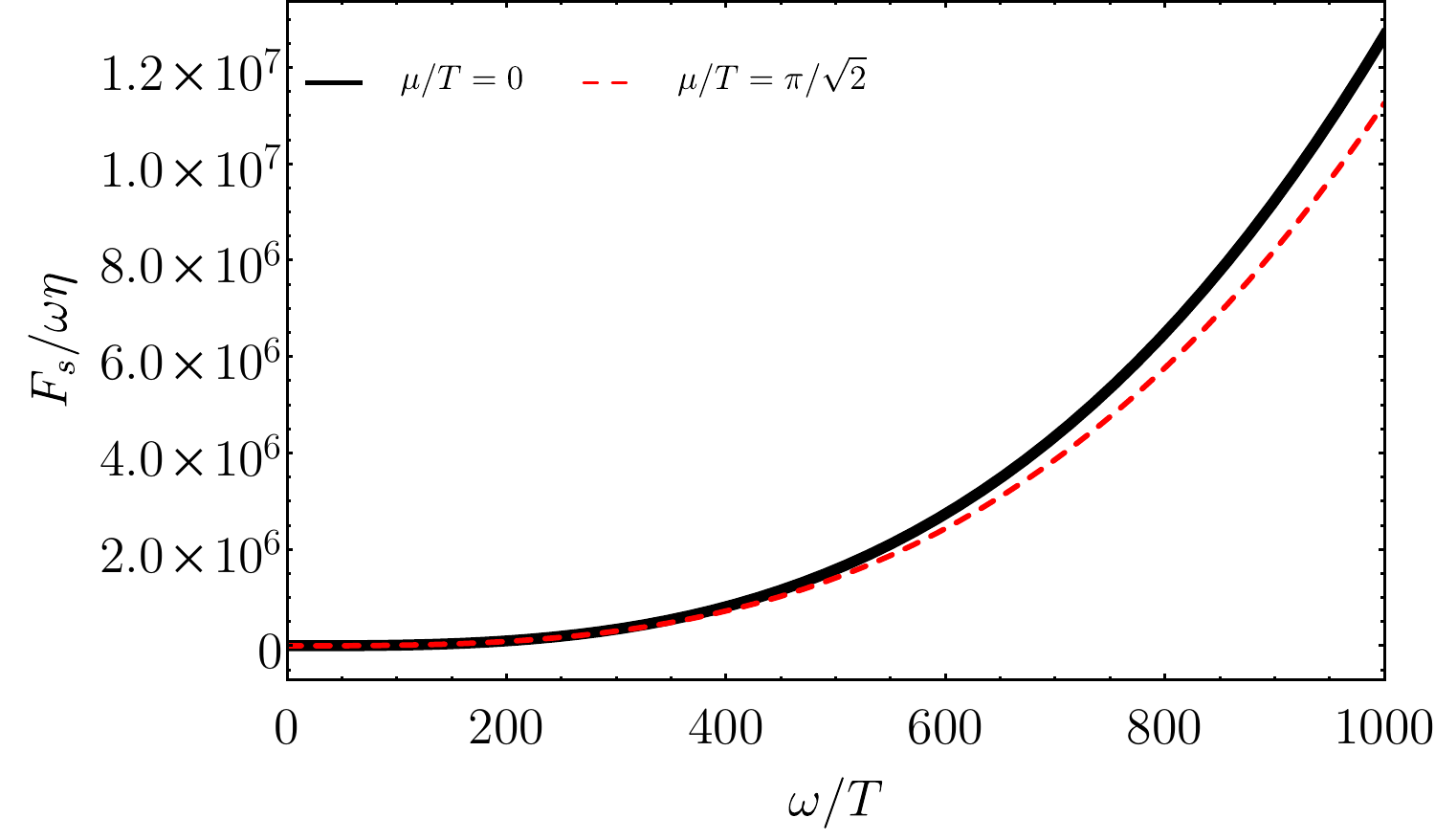} 
\caption{Normalized spectral function as a function of $\omega/T$ for $\mu/T = 0$ and $\mu/T = \pi/\sqrt 2$ (critical point). Both curves scale with $(\omega/T)^3$ as $\omega/T\to\infty$.}
\label{fig:spectralabs}
\end{figure}

We introduce a bulk response function \cite{Iqbal:2008by},\footnote{This response function should not be confused with the bulk viscosity, which is always zero in the conformal theory considered in this paper.}
\begin{equation}
\zeta \equiv 2\kappa_5^2 \dfrac{\Pi}{\omega\tilde{\varphi}},
\end{equation}
which allows us to reduce the linear second order differential equation \eqref{eq:eomspectral1} to a first order nonlinear Riccati equation,
\begin{align}
\label{eq:flow}
\zeta'+\dfrac{\omega g_{rr}}{\sqrt{-g}}\left[\zeta^2+g_{xx}^3\left(1+\dfrac{g_{tt}}{g_{xx}}\,\dfrac{k^2}{\omega^2}\right)\right]=0.
\end{align}
By requiring regularity at the horizon, one obtains the following horizon condition needed to solve the first order flow equation above,
\begin{align}
\zeta(r=r_H,\omega,\vec{k})=\pm ig_{xx}(r_H)^{3/2},
\end{align}
where we choose the positive sign, which corresponds to the in-falling wave at the horizon. From the membrane paradigm \cite{Iqbal:2008by}, assuming that the scalar disturbance corresponds to the $\delta g_x^y$ fluctuation of the metric, one recognizes this result as the shear viscosity
\begin{align}
\eta=\lim\limits_{\omega\to 0}\lim\limits_{\vec{k}\to 0}\dfrac{\textrm{Im}\,\zeta(r_H,\omega,\vec{k})}{2\kappa_5^2}.
\end{align}
Then, one may write down the following dimensionless ratio,
\begin{align}\label{eq:spectral_normalized}
\dfrac{F_s}{\omega\eta}=\dfrac{\textrm{Im}\,\zeta(r\to\infty,\omega,\vec{k})}{g_{xx}(r_H)^{3/2}}.
\end{align}
One can now numerically integrate Eq.\ \eqref{eq:flow} and then use Eq.\ \eqref{eq:spectral_normalized} to obtain the normalized spectral function. The results are shown in Fig.\ \ref{fig:spectralabs} for $\mu/T = 0$ (\AdS{5}-Schwarzschild) and $\mu/T = \pi/\sqrt 2$ (critical point), both evaluated at $k = 0$. We see that, naively, an increase in $\mu/T$ seems to have only a small effect on the spectral function, even as one approaches the critical point. However, this is due to the fact that in the ultraviolet limit, $\omega/T \to \infty$, the dimensionless ratio $F_s/\omega \eta$ scales as $(\omega/T)^3$, which overwhelms any poles or fluctuations in the plot for the spectral function.

\begin{figure}[t]
\centering
\begin{subfigure}{0.47\textwidth}
\includegraphics[width=\textwidth]{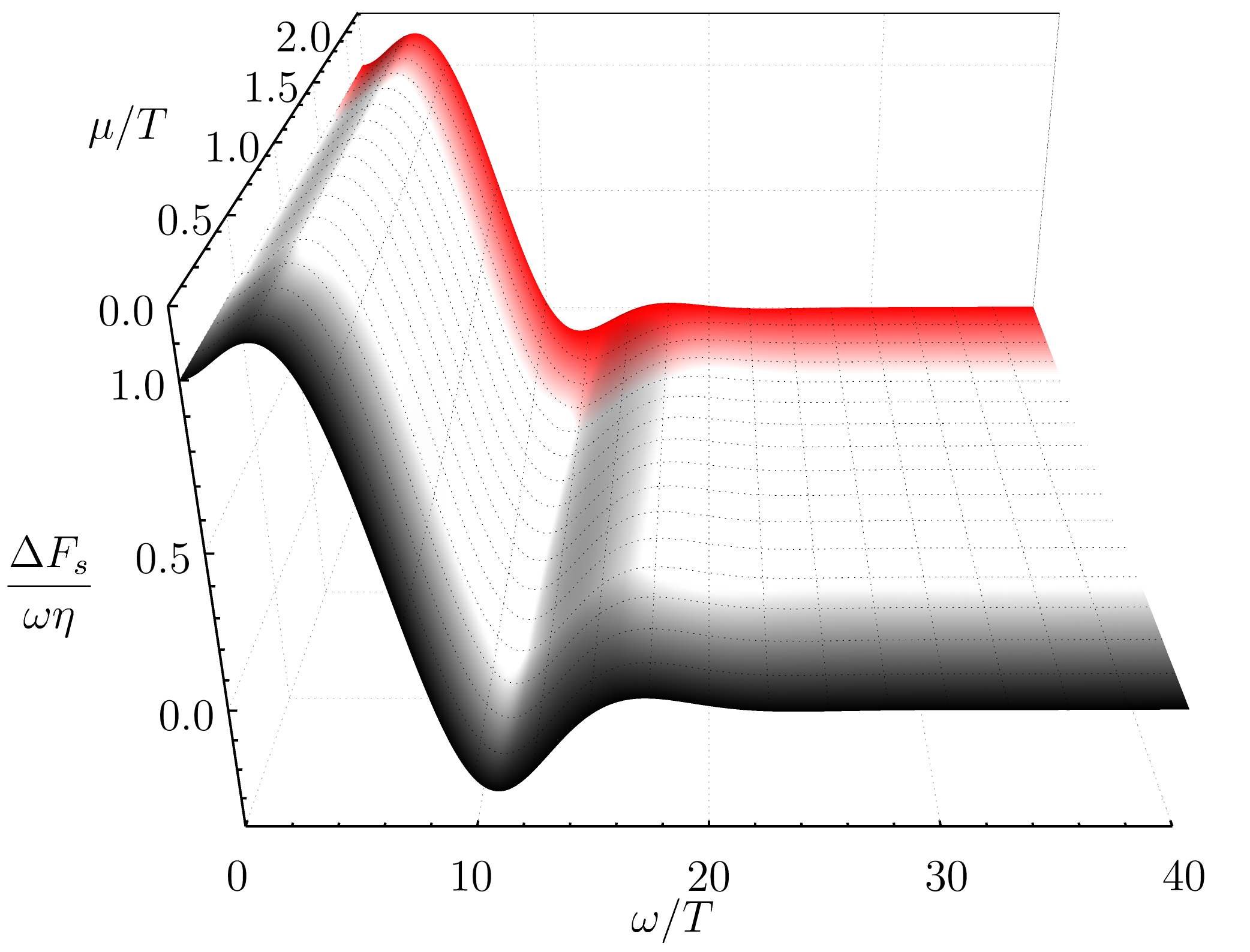}
\caption{}
\end{subfigure}
\begin{subfigure}{0.51\textwidth}
\includegraphics[width=\textwidth]{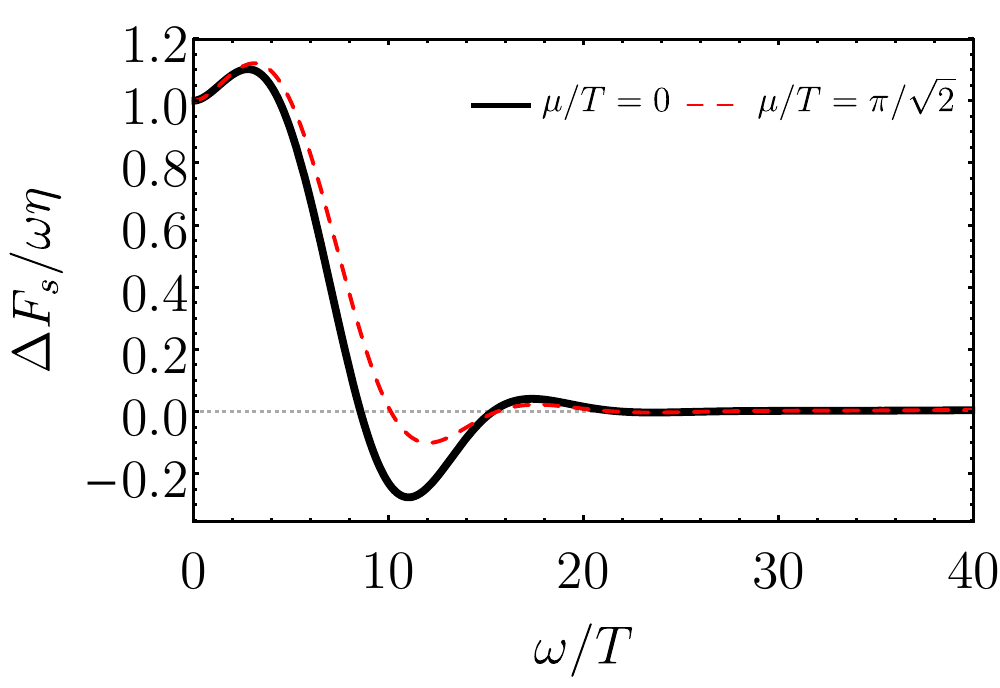}
\caption{}
\end{subfigure}
\caption{(Color online) (a) Full surface profile of the subtracted normalized spectral function as a function of $\omega/T$ and $\mu/T$, in the long wavelength limit, $k/T=0$. (b) Details of $\Delta F_s/\omega\eta$ for $\mu/T = 0$ and $\mu/T = \pi/\sqrt 2$ (critical point).}
\label{fig:spectralsub}
\end{figure}

\begin{figure}[t]
\centering
\includegraphics[width=0.6\textwidth]{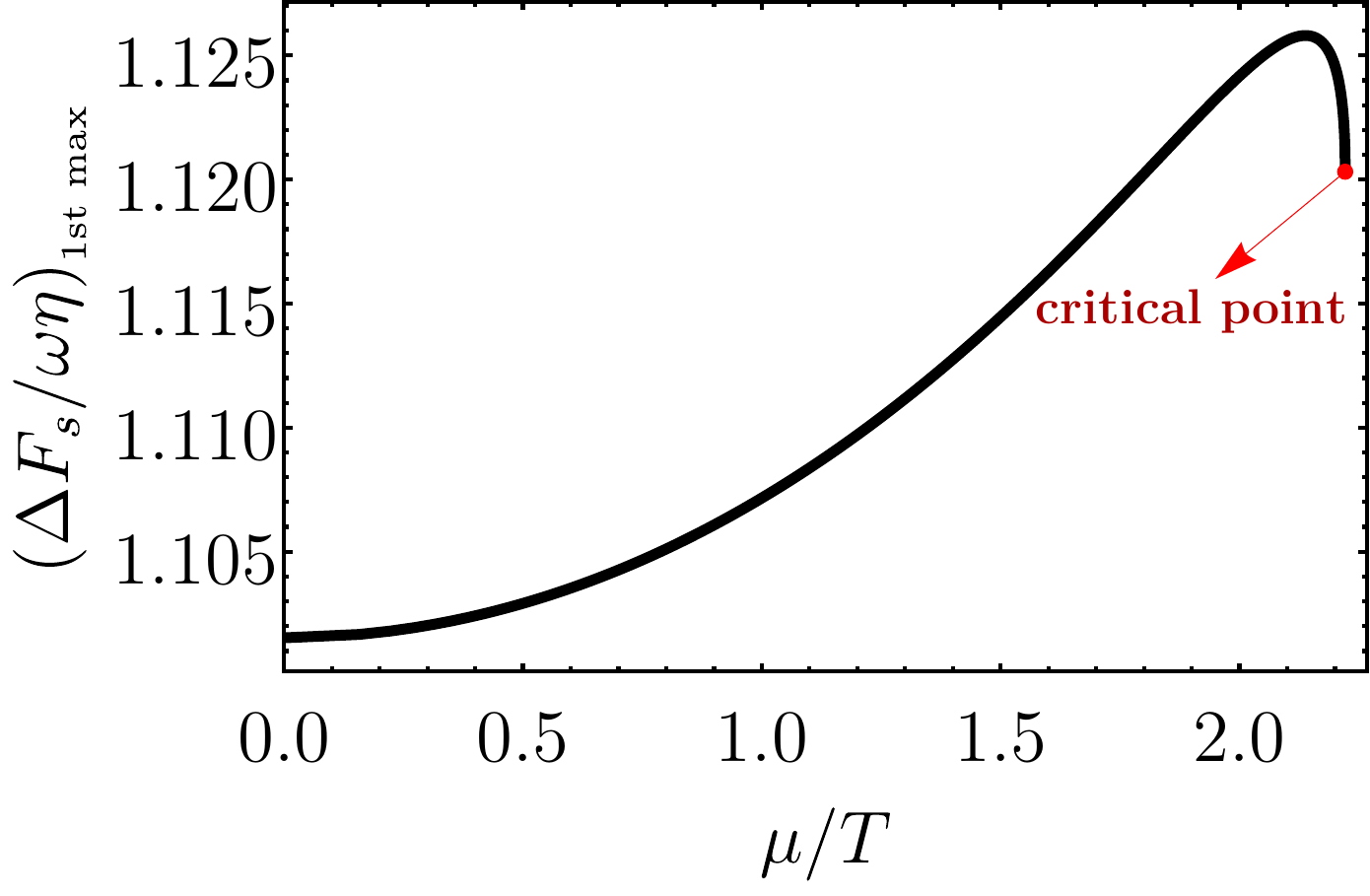}
\caption{(Color online) Height of the first peak of $\Delta F_s/\omega\eta$ as a function of $\mu/T$ in the long wavelength limit, $k/T=0$.}
\label{fig:spectralmax}
\end{figure}

Let us now consider a subtraction scheme which removes the scaling $(\omega/T)^3$ from the spectral function at some ultraviolet cutoff \cite{CaronHuot:2006te,Teaney:2006nc,Gursoy:2012bt},
\begin{align}
\frac{\Delta F_s}{\omega \eta} \equiv \frac{F_s(\omega,k=0)}{\omega \eta} - a \left( \frac{\omega}{T} \right)^3,
\end{align}
where $a$ is a constant defined by the asymptotic behavior of the spectral function, 
\begin{equation}
a \equiv \lim\limits_{\omega/T\to\infty}\left.\dfrac{F_s/\omega\eta}{(\omega/T)^3}\right|_{k=0}.
\end{equation}
We remark that in order to reliably perform the subtraction above, the flow equation \eqref{eq:flow} must be solved with high numerical accuracy. This becomes more difficult for larger values of the ultraviolet cutoff in $\omega/T$, which also requires one to increase the ultraviolet cutoff used to numerically parametrize the boundary in the radial coordinate. For the numerical evaluation of the first oscillations of the subtracted spectral function, one can safely take as the ultraviolet cutoff in the dimensionless frequency a value around $\omega/T \sim 40$.

In Fig.\ \ref{fig:spectralsub} we display the behavior of $\Delta F_s/\omega\eta$ as a function of $\mu/T$ and $\omega/T$ at $k/T = 0$, while in Fig.\ \ref{fig:spectralmax} we analyze the evolution of the height of its first peak as a function of $\mu/T$, which acquires an infinite slope at the critical point. By computing its derivative and fitting the numerical result close to the critical region using the same functional dependence as in Eq.\ \eqref{eq:fit}, one again obtains a critical exponent compatible with $1/2$. Thus, we find that the critical behavior found in the QNM's can also be found, albeit in an indirect manner, in the spectral function.

\section{Numerical procedures}
\label{sec:numerics}

\hspace{5 mm} The main numerical algorithm we employ in this work to find the QNM's is the pseudospectral method \cite{boyd01}. In this Appendix, we briefly review the main steps required to tackle the problem. The main advantage of the pseudospectral method, when compared to other methods used in the literature to calculate QNM's, is the ease of numerical implementation and the accuracy --- in general, it requires a modest number of collocation points and basis functions in order to compute several QNM's with high accuracy. The main disadvantage is the requirement of high numerical precision in the intermediate calculations, which still brings a drawback in terms of running time, since typical machine precision calculations result in spurious results for any but the lowest QNM's.

\subsection{Brief overview of the pseudospectral method}

\hspace{5 mm} The general spectral method, of which the pseudospectral (or collocant) method is a particular case,\footnote{The main difference between spectral and pseudospectral methods regards the determination of the coefficients $a_i$ to be specified in the sequel. As explained in \cite{boyd01}, the nomenclature used in the literature is a bit messy: both spectral and pseudospectral methods are known as spectral methods in a broad sense, due to the fact that they use a complete set of orthogonal functions. In the restricted sense, spectral methods (also called non-collocant methods) determine the generalized Fourier coefficients by exploiting the orthogonality of the basis, projecting down the unknown function $f(u)$. On the other hand, pseudospectral methods (also known as collocant methods) use a selected set of points on the function domain in order to build an interpolating polynomial. For the purposes of the present work, we will consider only pseudospectral (or collocant methods) in this restricted sense.} aims to solve the following general non-homogeneous differential equation for a complex function $f(u)$,
\begin{equation}
\label{eq:geneig}
\hat{L} f(u) = h(u),
\end{equation}
where $\hat{L}$ is a general differential operator and $h(u)$ is the non-homogeneous term. One may consider, for instance, the basic interval $u\in[0,1]$, which was used to define the radial holographic coordinate $u$ in the main text (with the boundary at $u=0$ and the horizon at $u=1$). In finite difference methods one discretizes the basic interval using a finite grid and introduce finite difference approximations for the derivatives. In both spectral and pseudospectral methods, one instead introduces a subset $\{\phi_i (u)\}^N_{i=0}$ of a complete set  $\{\phi_i(u)\}^{\infty}_{i=0}$ of orthogonal basis functions defined on the basic interval, approximating $f(u)$ by its truncated base expansion $f_N(u)$,
\begin{equation}
f(u) \approx f_N(u) = \sum\limits_{i=0}^N a_i \phi_i(u).
\end{equation}
Now we consider the pseudospectral method in order to determine the coefficients $a_i$ that best provide an approximation $f_N(u)$ for $f(u)$. First, one introduces the residual function $R\left(u;\{a_i\}\right)$ defined by,
\begin{equation}
R\left(u;\{a_i\}\right) = \hat{L} f_N(u) - h(u) = \sum\limits_{i=0}^N a_i \hat{L} \phi_i(u) - h(u).
\end{equation}
The strategy is to suitably choose a set of points $\{u_j\}_{j=0}^N$ (the so-called collocation points) on the domain of the basis functions $\phi_i$, and then fix $a_i$ such that the residual function is zero at $u_j$, that is, $R(u_j;\{a_i\}) = 0$. This generates a system of $N$ linear differential equations with $N$ variables, $\{a_i\}$,
\begin{equation}
\sum\limits_{i=0}^N a_i[\hat{L} \phi_i](u_j) = h(u_j)
\end{equation}
By solving for $\{a_i\}$ one determines the approximate solution $f_N(u)$.

There are several possible choices for the basis functions $\phi_i$ and collocation points $u_j$, depending on the symmetries of the problem and boundary conditions. A generally proposed basis defined in the interval $x\in[-1,1]$ is given by the Chebyshev polynomials, $ T_n(x) $.\footnote{The Chebyshev polynomials are defined to be the set of polynomials satisfying $ T_n(\cos\theta)=\cos(n\theta) $ for every angle $ \theta\in[0,\pi] $ and $ n\in\mathbb{N} $.} Since $ u\in[0,1] $, we can relate these variables by $ x=2u-1 $, then $\phi_i (u) = T_i(2u-1) $. A useful accompanying set of collocant points is given by the so-called Gauss-Lobatto grid, which for $u\in [0,1]$ can be written as,
\begin{equation}
u_j = \frac{1}{2}\left[1- \cos \left(\frac{j \pi}{N}\right)\right],\text{ for }j=0,1,\,..., N.
\end{equation}

Finally, we can solve the Generalized Eigenvalue Problem (GEP) which we are interested in by extending the operator $\hat{L}$ to include a dependence on a parameter $\lambda$, $\hat{L} \to \hat{L}(\lambda)=\hat{L}_0+\lambda\hat{L}_1$, and then taking $h(u)=0$, such as to search for eigenvalues $\lambda\equiv \omega/T$ satisfying $\hat{L}(\lambda) f(u) = 0$ with boundary values $ f(0)=f_0 $ and $ f(1)=f_1 $. The resulting matrix GEP is then,
\begin{align}
\left(A_0+A_1\lambda\right)a=0,
\end{align}
where $ a=\{a_i\} $ and $ A_k=\{A_{k,ij}\}=\{[\hat{L}_k\phi_i](u_j)\} $ with $ k\in\{0,1\} $.

\subsection{Implementation details}

\hspace{5 mm} We used a Chebyshev basis with $N$ basis functions and collocant points, and then employed the Arnoldi method (via the built-in \texttt{Eigensystem[\,]} procedure in \emph{Wolfram's Mathematica} \cite{math}) to solve the resulting GEP for $\omega/T$.\footnote{In the particular case of the \AdS{5}-Schwarzschild background, a didactic sample calculation of the QNM spectra using the pseudospectral method can be found in L.~Yaffe's notebook for the 2014 Mathematica Summer School on Theoretical Physics \cite{yaffenb}.} For higher order QNM's, one needs to use high numerical precision from the outset. Denoting the number of floating-point digits used in the calculations by $M$, the final error estimate of the numerical QNM's is mainly controlled by the number of basis functions $N$ and the numerical precision parameter $M$. We have checked that using $M = 60$ and $N = 80$ yielded the 20 first QNM's with good accuracy --- this was the setting used in most of the paper. For higher order QNM's, such as the ones shown in Figs.\ \ref{fig:qnmtyp} and \ref{fig:qnmtypcond}, we used $M=N=100$. We verified the stability of the QNM spectra involving the desired modes by doubling the number of basis points $N$ or the numerical precision parameter $M$, following the error control procedure discussed in \cite{boyd01}.

\subsection{Calculation of the critical exponents}
\label{sec:error}

\hspace{5 mm} Let us now discuss the numerical procedure we followed in order to determine the critical exponent $\theta$ of the divergent quantities near the critical point $\mu/T = \pi/\sqrt 2$, by performing fits of the asymptotic form \eqref{eq:fit} as a function of $\mu/T$.

\begin{table}
\centering
\begin{tabular}{|ccc|} \hline
    {Interval} & {Starting $d(T \tau_\mathrm{eq})/d(\mu/T)$} & {Ending $d(T \tau_\mathrm{eq})/d(\mu/T)$}  \\ \hline \hline
    1  & 0.05 & 0.10 \\
    2  & 0.10  & 0.15 \\
    3  & 0.15  & 0.20 \\
    4  & 0.20  & 0.25 \\ \hline
    5  & 0.25   & 0.30 \\
    6  & 0.30  & 0.50 \\
    7  & 0.50  & 1.0 \\
    8  & 1.0  & 5.0 \\ \hline
    9  & 5.0  & 15.0 \\
    10 & 15.0  & 40.0 \\
    11 & 40.0  & 70.0 \\
    12 & 70.0  & 150.0 \\ \hline
	13 & 150.0 & 250.0 \\ 
	14 & 250.0 & 530.0 \\ \hline
\end{tabular}
\caption{\label{tab:intervals}$d(T \tau_\mathrm{eq})/d(\mu/T)$ intervals used for the fit procedure.}
\end{table}

\begin{figure}[t]
\centering
\includegraphics[width=0.6\textwidth]{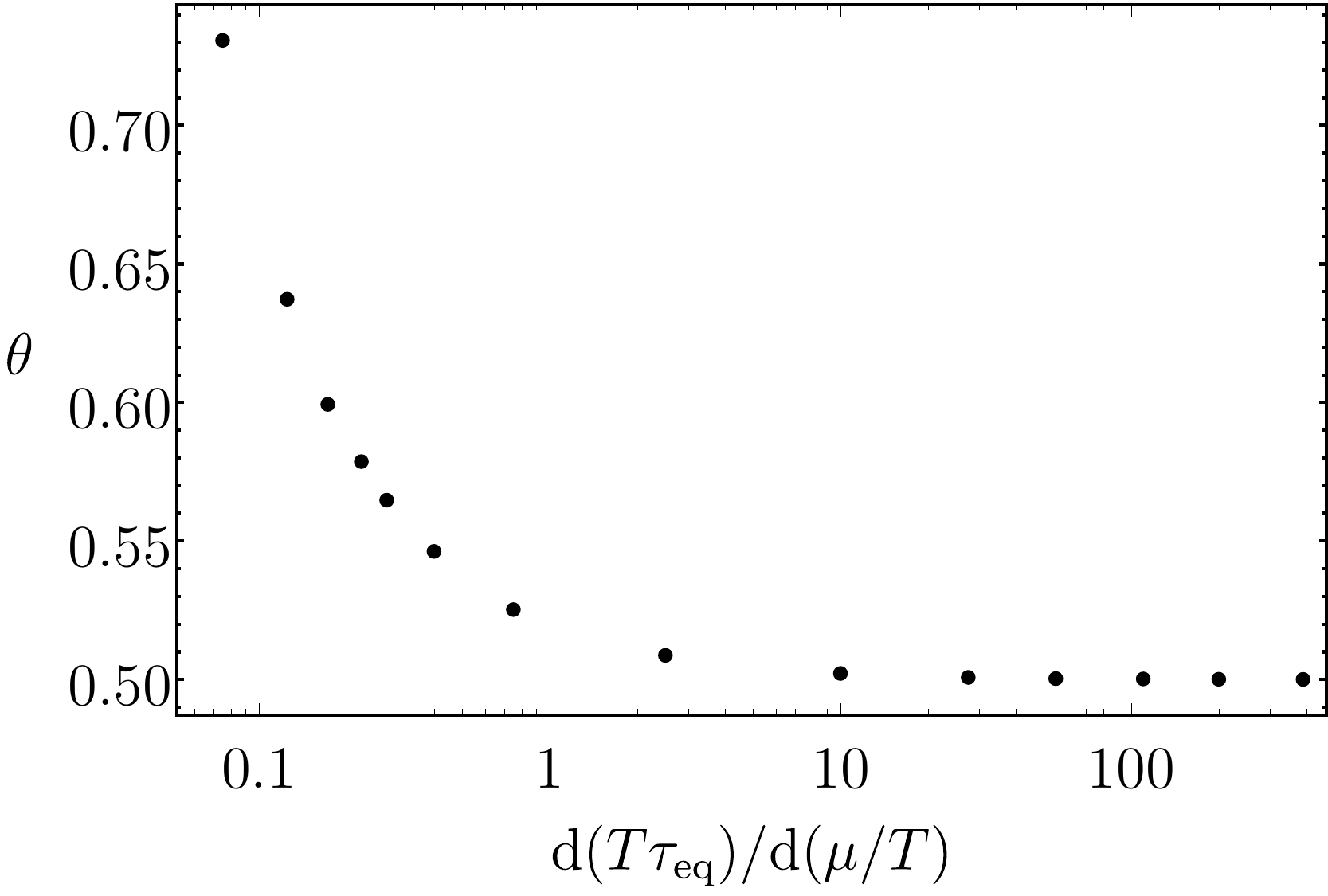}
\caption{Fit results for the critical exponent $\theta$ for each subinterval in $d(T \tau_\mathrm{eq})/d(\mu/T)$. The abscissas were chosen at the midpoint of the corresponding subinterval.}
\label{fig:fit}
\end{figure}

We used a first order central difference formula with step size $h \equiv \Delta (\mu/T)$ in order to compute the required numerical derivatives. By varying $h$ from $10^{-5}$ to $10^{-11}$, and taking into account that we have an accuracy of several digits in the computed observables, we estimated the error introduced by the numerical differentiation to be of the order of $10^{-7}$.

Taking as a specific example the calculation of the critical exponent for the equilibration time in the external scalar channel, in order to check that the asymptotic form \eqref{eq:fit} is valid close to the critical point $\mu/T = \pi/\sqrt 2$, we split $d(T \tau_\mathrm{eq})/d(\mu/T)$ into 14 subintervals and then performed a least squares fit of Eq.\ \eqref{eq:fit} to the resulting data within each subinterval, with each of them populated by 100 points evenly spaced in $\mu/T$. In Table \ref{tab:intervals} we specify the subintervals used for the determination of the critical exponent of $d(T \tau_\mathrm{eq})/d(\mu/T)$ in the external scalar channel. In Fig.\ \ref{fig:fit} we display the convergence of the fitted critical exponent $\theta$ to the value $1/2$. The estimate of the least squares standard error in the value of $\theta$ is of the order of $10^{-7}$ for the last interval.

\bibliographystyle{JHEP}
\bibliography{Bibliography}

\end{document}